\documentclass[pre,tighten]{revtex4}

\usepackage{natbib}
\usepackage{amssymb,amsmath}
\usepackage{epsfig}
\usepackage{bm}
\usepackage{color}
\usepackage{graphicx}
\usepackage{hyperref}

\usepackage{graphicx}

\usepackage{color}
\newcommand{\vicente}[1]{{ #1}}

\newcommand\beq{\begin{equation}}
\newcommand\eeq{\end{equation}}
\newcommand\beqa{\begin{eqnarray}}
\newcommand\eeqa{\end{eqnarray}}
\newcommand{\dd}{\text{d}}

\newcommand{\al}{\alpha}

\begin{document}

\title{Simple shear flow in granular suspensions: Inelastic Maxwell models and BGK-type kinetic model}

\author{Rub\'en G\'omez Gonz\'alez\footnote[1]{Electronic address: ruben@unex.es}}
\affiliation{Departamento de F\'{\i}sica,
Universidad de Extremadura, E-06071 Badajoz, Spain}
\author{Vicente Garz\'{o}\footnote[2]{Electronic address: vicenteg@unex.es;
URL: http://www.unex.es/eweb/fisteor/vicente/}}
\affiliation{Departamento de F\'{\i}sica and Instituto de Computaci\'on Cient\'{\i}fica Avanzada (ICCAEx), Universidad de Extremadura, E-06071 Badajoz, Spain}

\begin{abstract}
The Boltzmann kinetic equation for low-density granular suspensions under simple shear flow is considered to determine the velocity moments through the fourth degree. The influence of the interstitial gas on solid particles is modeled by a viscous drag force term plus a stochastic Langevin-like term. Two independent but complementary approaches are followed to achieve exact results. First, to keep the structure of the Boltzmann collision operator, the so-called inelastic Maxwell models (IMM) are considered. In this model, since the collision rate is independent of the relative velocity of the two colliding particles, the forms of the collisional moments can be obtained without the knowledge of the velocity distribution function. As a complement of the previous effort, a BGK-type kinetic model adapted to granular gases is solved to get the velocity moments of the velocity distribution function.  The analytical predictions of the rheological properties (which are \emph{exactly} obtained in terms of the coefficient of restitution $\al$ and the reduced shear rate $a^*$) show in general an excellent agreement with event-driven simulations performed for inelastic hard spheres. In particular, both theoretical approaches show clearly that the temperature and non-Newtonian viscosity exhibit an $S$ shape in a plane of stress-strain rate (discontinuous shear thickening effect). With respect to the fourth-degree velocity moments, we find that while those moments have unphysical values for IMM in a certain region of the parameter space of the system, they are well defined functions of both $\al$ and $a^*$ in the case of the BGK kinetic model. The explicit shear-rate dependence of the fourth-degree moments beyond this critical region is also obtained and compared against available computer simulations.
\end{abstract}

\draft
\date{\today}
\maketitle

\section{Introduction}
\label{sec1}

One of the most challenging problems in non-Newtonian gas-solid suspensions is the so-called discontinuous shear thickening, namely, when the non-Newtonian shear viscosity of the suspension drastically increases with increasing the shear rate. This problem \vicente{(which mainly occurs in concentrated suspensions of particles such as mixtures of cornstarch in water \cite{BJ14})} has attracted the attention of physicists \cite{B89,LDHH05,BJ09,MW11,CPNC11,OH11,H13,SMMD13,BJ14,KFZFS18} in the last few years as a typical nonequilibrium discontinuous transition between a liquid-like phase and a solid-like phase. \vicente{As pointed out by Brown and Jaeger \cite{BJ09}, there are essentially three different possible mechanisms  to explain this dramatic version of shear thickening. One mechanism is hydroclustering where the particles tend to move together into clusters under shear and hence, lubrication drag forces between particles are increased due to this type of rearrangement \cite{BB85,WB09}. A second mechanism \cite{H74,H82} is related to a transition in the microstructure from ordered layers at small shear rates to disordered layers at higher shear rates (order-disorder transition). Finally, a third mechanism is dilatancy in which the packing volume of particles dilates (expands) with increasing the shear rate \cite{CHH05,BJ12}.}

Although \vicente{most of the studies on shear thickening have been focused on very dense systems, it would be convenient to analyze} relatively low-density systems where kinetic theory tools conveniently adapted to account for the dissipative character of collisions can be employed to unveil in a clean way the microscopic mechanisms involved in the discontinuous shear thickening. In particular, some previous papers \cite{TK95,SMTK96,SA17} demonstrated the existence of a nonequilibrium discontinuous transition for the granular temperature between a \emph{quenched} state (a low-temperature state) and an \emph{ignited} state (a high-temperature state) in a granular suspension under simple shear flow described by the Boltzmann equation.

A more recent work has been performed by Hayakawa \emph{et al.} \cite{HTG17} in the context of the Enskog kinetic equation for a moderately dense gas-solid suspension under simple shear flow. In contrast to the previous attempts \cite{TK95,SMTK96,SA17}, the effect of the interstitial gas on solid particles is modeled via a viscous drag force plus a stochastic Langevin-like term. The Enskog equation is solved by means two complementary routes: (i) Grad's moment method and (ii) event-driven Langevin simulations for inelastic hard spheres (IHS). Both approaches clearly show a transition from the discontinuous shear thickening (observed for very dilute gases) to the continuous shear thickening as the density of the system increases.

On the other hand, as in the case of elastic collisions \cite{C88,CC70,FK72}, a limitation of the theoretical results obtained in Ref.\ \cite{HTG17} is that they were \emph{approximately} obtained by means of Grad's moment method (namely, by considering the leading terms in a Sonine polynomial expansion of the velocity distribution function). The source of this limitation comes mainly from the form of the collision rate for hard spheres (which is proportional to the magnitude of the normal component of the relative velocity of the two colliding spheres) appearing inside the Boltzmann collision operator. As for elastic collisions, the lack of exact analytical results of the Boltzmann equation has stimulated the use of the so-called inelastic Maxwell models (IMM), where the collision rate is independent of the relative velocity. IMM have received a lot of attention in the last few years since they allow to assess the influence of inelasticity on the dynamic properties of the system without introducing additional approximations.

Another possible way of overcoming the mathematical difficulties of the Boltzmann collision operator is to consider a kinetic model. The kinetic models retain the relevant physical properties of the Boltzmann kinetic equation and are more tractable than the true kinetic equation. This kind of approach has been widely employed in the case of dilute gases with elastic collisions \cite{GS03}, where several exact solutions in far from equilibrium states have been obtained in the past and shown to be in good agreement with numerical solutions of the Boltzmann equation. Here, we will consider a Bhatnagar--Gross--Krook (BGK) model kinetic equation \cite{BDS99} for granular suspensions to complement the theoretical results derived from the Boltzmann equation for IMM.

The objective of this paper is to determine the dynamic properties of a granular suspension under simple or uniform shear flow (USF). This state is characterized by a constant density, a uniform granular temperature, and a linear velocity profile $U_x=ay$, where $a$ is the constant shear rate. We are interested here in the steady state where the system admits a non-Newtonian hydrodynamic description characterized by shear-rate dependent viscosity and normal stress differences. The evaluation of the rheological properties is one of the most important goals of the present contribution. However, although these transport properties (which are related with the second-degree velocity moments) are physically important, higher degree velocity moments offer also an important piece of information about the velocity distribution function, especially in the high velocity region. By symmetry reasons, the third-degree moments vanish in the steady state in the USF problem. Thus, beyond the rheological properties, the first nontrivial moments are the fourth-degree moments. Their knowledge allows us to gauge partially the joint effect of shearing, interstitial gas, and inelasticity on the velocity distribution function.

The efforts of computing the second- and fourth-degree moments for IMM in the USF problem may be justified at least for three different reasons. First, the determination of the rheological properties can allow us to assess the degree of reliability of IMM to capture the main trends observed previously in sheared granular suspensions of IHS. As a second reason, it is interesting to explore whether or not the divergence of the fourth-degree moments for elastic \cite{SGBD93,SG95} and inelastic \cite{SG07} Maxwell gases beyond a certain critical shear rate is also present in granular suspensions and, if so, to what extent. Finally, the knowledge of the fourth-degree moments is needed to evaluate the relevant transport coefficients characterizing states close to the USF state \cite{G07bis}. This knowledge will allow us to analyze the stability of the (steady) USF state in granular suspensions.

The plan of the paper is as follows. In section \ref{sec2}, the Boltzmann equation for granular suspensions under USF is introduced and the corresponding balance equations for the densities of mass, momentum, and energy are deduced. Section \ref{sec3} deals with the calculations carried out for IMM for the second- and fourth-degree moments. Since the (scaled) granular temperature $\theta$ is a multi-evaluated function of the (reduced) shear rate $a^*$, it is more convenient to analyze the divergence of the fourth-degree moments taking $\theta$ as input parameter instead of $a^*$. Therefore, in a way similar to the case of elastic Maxwell molecules \cite{SGBD93,SG95} and \emph{dry} granular gases (namely, when the effect of gas phase on solid particles is neglected) \cite{SG07}, we find that, for a given value of $\al$, those moments tend to infinity for certain critical values $\theta_c^{(1)}$ and $\theta_c^{(2)}$ of the granular temperature. More specifically, those moments have unphysical values in the region $\theta_c^{(1)}<\theta<\theta_c^{(2)}$. The results derived from the BGK kinetic model are displayed in section \ref{sec4} where it is shown first that the BGK predictions of the rheological properties coincide with those obtained by solving the Boltzmann equation by means of Grad's moment method \cite{HTG17}. In addition and in contrast with IMM, the BGK moments are well defined functions in the complete parameter space of the system. Comparison between theory and computer simulations at the level of the rheological properties is performed in section \ref{sec5}. The excellent agreement found here among the different tools confirms again the reliability of both theoretical approaches (Boltzmann equation for IMM and BGK model for IHS) for studying non-Newtonian transport properties in sheared granular suspensions. Finally, the paper is closed in section \ref{sec6} with some concluding remarks.

\section{Boltzmann kinetic equation for sheared granular suspensions}
\label{sec2}

\subsection{Boltzmann kinetic equation for granular suspensions}

Let us consider a set of solid particles of diameter $\sigma$ and mass $m$ immersed in a viscous gas. Since the grains which make up a granular material are of a macroscopic size, their collisions are inelastic. In the simplest model, the inelasticity of collisions is characterized by a (positive) constant coefficient of normal restitution $\al \leq 1$, where $\al=1$ corresponds to elastic collisions (ordinary gases). In the low-density regime, the one-particle velocity distribution function of solid particles  $f(\mathbf{r}, \mathbf{v}; t)$ obeys the Boltzmann kinetic equation
\beq
\label{2.1}
\frac{\partial f}{\partial t}+\mathbf{v}\cdot \nabla f+\mathcal{F} f=J[\mathbf{v}|f,f],
\eeq
where $J[f,f]$ is the Boltzmann collision operator \cite{BP04} and $\mathcal{F}$ is an operator representing the fluid-solid
interaction force that models the effect of the viscous gas on solid particles. In order to fully account for the influence of
the interstitial molecular fluid on the dynamics of grains, a instantaneous fluid force model is employed \cite{GTSH12,GFHY16,HTG17}. For low Reynolds numbers, it is assumed that the external force $\mathbf{F}$ acting on solid particles is composed by two independent terms. One term corresponds to a viscous drag force $\mathbf{F}^{\text{drag}}$ proportional to the (instantaneous) velocity of particle $\mathbf{v}$. This term takes into account the friction of grains on the viscous gas. Since the model attempts to mimic gas-solid flows, the drag force is defined in terms of the relative velocity $\mathbf{v}-\mathbf{U}_g$ where $\mathbf{U}_g$ is the (known) mean flow velocity of the surrounding molecular gas. Thus, the drag force is defined as
\beq
\label{2.2}
\mathbf{F}^{\text{drag}}=-m\gamma \left(\mathbf{v}-\mathbf{U}_g\right),
\eeq
where $\gamma$ is the drag or friction coefficient. The second term in the total force corresponds to a stochastic force that tries to simulate the kinetic energy gain due to eventual collisions with the (more rapid) molecules of the background fluid. It does this by adding a random velocity to each particle between successive collisions \cite{WM96}. This stochastic force $\mathbf{F}^{\text{st}}$ has the form of a Gaussian white noise with the properties \cite{K81}
\beq
\label{2.3}
\langle \mathbf{F}_i^{\text{st}}(t) \rangle =\mathbf{0}, \quad
\langle \mathbf{F}_i^{\text{st}}(t) \mathbf{F}_j^{\text{st}}(t') \rangle = 2 m^2 \gamma T_\text{ex} \mathsf{I} \delta_{ij}\delta(t-t'),
\eeq
where $\mathsf{I}$ is the unit tensor and $i$ and $j$ refer to two different particles. Here, $T_\text{ex}$ can be interpreted as the temperature of the background (or bath) fluid. In the context of the Boltzmann equation, the stochastic external force is represented by a Fokker--Planck operator of the form $\mathcal{F}^{\text{st}}f\to -(\gamma T_\text{ex}/m)\partial^2 f/\partial v^2$ \cite{K81,NE98}. Note that the strength of correlation in Eq.\ \eqref{2.3} has been chosen to be consistent with the fluctuation-dissipation theorem for elastic collisions \cite{K81}. In addition, although the drift coefficient $\gamma$ is in general a tensor, in the case of very dilute suspensions it may be assumed to be an scalar proportional to the square root of $T_\text{ex}$ because the drag coefficient is proportional to the viscosity of the solvent \cite{KH01}.

Therefore, according to Eqs. \eqref{2.2} and \eqref{2.3}, the forcing term $\mathcal{F}f$ can be written as
\beq
\label{2.3.1}
\mathcal{F}f=-\gamma\Delta \mathbf{U}\frac{\partial f}{\partial \mathbf{v}}-\gamma\frac{\partial}{\partial \mathbf{v}}\cdot \mathbf{V} f-\gamma \frac{T_{\textup{ex}}}{m}\frac{\partial^2 f}{\partial v^2},
\eeq
and the Boltzmann equation \eqref{2.1} reads
\beq
\label{2.4}
\frac{\partial f}{\partial t}+\mathbf{v}\cdot \nabla f-\gamma\Delta \mathbf{U}\frac{\partial f}{\partial \mathbf{v}}-\gamma\frac{\partial}{\partial \mathbf{v}}\cdot \mathbf{V} f-\gamma \frac{T_{\textup{ex}}}{m}\frac{\partial^2 f}{\partial v^2}=J[\mathbf{V}|f,f].
\eeq
Here, $\Delta \mathbf{U}=\mathbf{U}-\mathbf{U}_g$, $\mathbf{V}=\mathbf{v}-\mathbf{U}$ is the peculiar velocity,
\beq
\label{2.5}
\mathbf{U}(\mathbf{r},t)=\frac{1}{n(\mathbf{r},t)}\int \dd\mathbf{v}\; \mathbf{v} f(\mathbf{r},\mathbf{v},t)
\eeq
is the mean particle velocity, and
\beq
\label{2.6}
n(\mathbf{r},t)=\int \dd\mathbf{v}\; f(\mathbf{r},\mathbf{v},t)
\eeq
is the number density. Another relevant hydrodynamic field is the \emph{granular} temperature $T(\mathbf{r},t)$ defined as
\beq
\label{2.7}
T(\mathbf{r},t)=\frac{m}{d n(\mathbf{r},t)} \int \dd\mathbf{v}\; V^2 f(\mathbf{r},\mathbf{v},t).
\eeq
The suspension model \eqref{2.4} is a simplified version of the model proposed in Ref.\ \cite{GTSH12} for monodisperse gas-solid flows at moderate density. In this latter model, the friction coefficient of the drag force and the strength of the correlation are considered to be different. Here, both coefficients are assumed to be the same for the sake of simplicity. Another relevant point of the model \eqref{2.3} is that the form of the Boltzmann collision operator $J[f,f]$ is assumed to be the same as for a dry granular gas (i.e., when the influence of the interstitial gas is neglected) and hence, the collision dynamics does not contain any parameter of the environmental gas. This means that while the inertia of particles is assumed to be relevant, the inertia of the gas phase is considered to be negligible. As has been previously discussed in several papers \cite{K90,TK95,SMTK96,KH01,WKL03}, the above assumption requires that the mean-free time between collisions is assumed to be much less than the time needed by the fluid forces to significantly affect the dynamics of solid particles. Thus, the suspension model \eqref{2.3} is expected to be reliable in situations where the gas phase has a weak impact on the motion of grains. This assumption fails for instance in the case of liquid flows (high density) where the stresses exerted by the background fluid on grains are expected to be important and hence, the presence of fluid should be accounted for in the collision process.

The Boltzmann collision operator conserves the mass and momentum but the energy is not conserved:
\beq
\label{2.8}
\int\; \dd\mathbf{v}J[\mathbf{v}|f,f]=0, \quad  \int\; \dd\mathbf{v}\; m \mathbf{v} J[\mathbf{v}|f,f]=\mathbf{0},
\eeq
\beq
\label{2.9}
\int\; \dd\mathbf{v}\; \frac{m}{2} V^2 J[\mathbf{v}|f,f]=-\frac{d}{2}n T \zeta,
\eeq
where $\zeta$ is the cooling rate due to inelastic collisions between the particles. From Eqs.\ \eqref{2.4}, \eqref{2.8}, and \eqref{2.9}, the macroscopic balance equations for the granular suspension can be obtained. They are given by
\begin{equation}
\label{2.10}
D_{t}n+n\nabla \cdot {\bf U}=0,
\end{equation}
\begin{equation}
\rho D_{t}\mathbf{U}+\nabla \cdot \mathsf{P}=-\rho \gamma \Delta \mathbf{U},
\label{2.11}
\end{equation}
\begin{equation}
D_{t}T+\frac{2}{dn} \left( \nabla \cdot {\bf q}+\mathsf{P}:\nabla {\bf U}\right) =
2 \gamma \left(T_\text{ex}-T\right)-\zeta \,T.
\label{2.12}
\end{equation}
Here, $D_t\equiv \partial_t+\mathbf{U}\cdot \nabla$, $\rho=m n$ is the mass density,
\beq
\label{2.13}
\mathsf{P}=\int\; \dd\mathbf{v}\; m\;\mathbf{V}\mathbf{V} f(\mathbf{v})
\eeq
is the pressure tensor, and
\beq
\label{2.13a}
\mathbf{q}=\int\; \dd\mathbf{v}\; \frac{m}{2}\; V^2\mathbf{V} f(\mathbf{v})
\eeq
is the heat flux.

To completely define the suspension model \eqref{2.4}, it still remains to explicitly write the form of the Boltzmann collision operator $J[f,f]$. The prototypical model of granular gases consists of a gas of IHS and hence, the collision rate appearing in the Boltzmann operator is proportional to the relative velocity of colliding spheres. Although this is an interaction model widely used in granular literature, it is generally not possible to get \emph{exact} analytical results from the Boltzmann equation for IHS, especially in far from equilibrium states such as the USF. As a consequence, most of the analytical results reported in the literature in the context of the Boltzmann equation for IHS have been obtained by introducing additional, and sometimes uncontrolled, approximations. In particular, the rheological properties of granular suspensions under USF have been recently determined \cite{HTG17} by means of Grad's moment method. Therefore, from a theoretically oriented point of view, if one desires to overcome the mathematical intricacies associated with the Boltzmann operator for IHS and derive exact results, one has at least two fruitful routes. One of them is to retain the mathematical structure of the Boltzmann equation but consider IMM. For this interaction model the collision rate is independent of the relative velocity of the colliding pair. This allows for a number of nice mathematical properties of the Boltzmann collision operator. The second possibility is to consider a kinetic model of the Boltzmann equation, namely, one replaces the operator $J[f,f]$ by a simpler collision model that otherwise retains the most relevant physical properties of the true Boltzmann collision operator. IMM will be considered in Sec.\ \ref{sec3} while the kinetic model will be employed in Sec.\ \ref{sec4}.

\subsection{Steady uniform shear flow}

Let us assume that the granular suspension is under USF. As said in the Introduction, this state is macroscopically defined by a constant density $n$, a spatially uniform temperature $T(t)$, and a flow velocity $U_i=a_{ij}r_j$, where $a_{ij}=a\delta_{ix}\delta_{jy}$, $a$ being the constant shear rate. In addition, as usual in uniform sheared suspensions \cite{TK95,SMTK96,SA17,ChVG15}, the average velocity of
particles follows the velocity of the fluid phase and so, $\mathbf{U}=\mathbf{U}_g$. One of the main advantages of the USF at a microscopic level is that in this state all the space dependence of the one-particle velocity distribution function $f(\mathbf{r}, \mathbf{v}, t)$ occurs through its dependence on the peculiar velocity $\mathbf{V}=\mathbf{v}-\mathbf{U}(\mathbf{r})$ \cite{DSBR86}. Thus, at a more fundamental level, the USF is defined as that which is spatially homogeneous when the velocities of particles are referred to a Lagrangian frame moving with the linear velocity field $U_i$. In this frame, the distribution function adopts the form
\beq
\label{2.14}
f(\mathbf{r},\mathbf{v};t)=f(\mathbf{V};t),
\eeq
and hence, in the steady state, the Boltzmann equation \eqref{2.4} reduces to
\beq
\label{2.15}
-aV_y\frac{\partial f}{\partial V_x}-\gamma\frac{\partial}{\partial \mathbf{V}}\cdot \mathbf{V} f-\gamma \frac{T_{\textup{ex}}}{m}\frac{\partial^2 f}{\partial V^2}=J[\mathbf{V}|f,f].
\eeq
Equation \eqref{2.15} is invariant under the transformations $(V_x, V_y)\to (-V_x, -V_y)$ and $V_j\to -V_j$ for $j\neq x,y$.

In the USF problem, the heat flux vanishes ($\mathbf{q}=\mathbf{0}$) and the (uniform) pressure tensor $\mathsf{P}$ is the relevant flux. Moreover, the conservation equations \eqref{2.10} and \eqref{2.11} hold trivially and in the steady state the balance equation \eqref{2.12} for the granular temperature becomes
\beq
\label{2.16}
-\frac{2}{dn}a P_{xy}-\zeta T+2\gamma\left(T_\text{ex}-T\right)=0.
\eeq
Equation \eqref{2.16} implies that in the steady state the viscous heating term ($-aP_{xy}>0$) plus the energy gained by grains due to collisions with the interstitial fluid ($\gamma T_\text{ex}$) is exactly compensated by the cooling terms arising from collisional dissipation ($\zeta T$) and viscous friction ($\gamma T$). Thus, for a given value of the environmental temperature $T_\text{ex}$, the (steady) scaled temperature $\theta\equiv T/T_\text{ex}$ is a function of the coefficient of restitution $\al$ and the (scaled) shear rate $a^*\equiv a/\gamma$. Of course, in the absence of shear flow ($a=0$), the solution to Eq.\ \eqref{2.16} is $T=T_\text{ex}$ for elastic collisions ($\al=1$ and so, $\zeta=0$) as expected. Note that in contrast to \emph{dry} granular gases ($\gamma=0$), a steady state is still possible for \emph{sheared} suspensions when the collisions between the solid particles are elastic.

The USF state is in general non-Newtonian. This can characterized by generalized transport coefficients measuring their departure from their corresponding Navier--Stokes forms. Thus, a non-Newtonian shear viscosity coefficient $\eta(\al,a)$ is defined as
\beq
\label{2.17}
\eta=-\frac{P_{xy}}{a}.
\eeq
Moreover, while in the Navier--Stokes domain $P_{xx}=P_{yy}=P_{zz}$, normal stress differences are expected in the USF state ($P_{xx}\neq P_{yy}\neq P_{zz}$). All the above properties may be easily identified from the knowledge of the (reduced) shear stress $P_{xy}^*$ and the (reduced) diagonal elements $P_{xx}^*$, $P_{yy}^*$, and $P_{zz}^*$, where
\beq
\label{2.17.1}
P_{ij}^*\equiv \frac{P_{ij}}{n T_\text{ex}}.
\eeq
It is quite apparent that the determination of the rheological properties requires to solve the Boltzmann equation \eqref{2.15}. As said before, Grad's moment method \cite{G49} has been used to solve Eq.\ \eqref{2.15} for IHS \cite{HTG17}. Grad's moment method is based on the expansion of the velocity distribution function in a complete set of orthogonal polynomials (generalized Hermite
polynomials), the coefficients being the corresponding velocity moments. However, given that
the (infinite) hierarchy of moment equations is not a closed set of equations, one has to truncate the
above expansion after a certain order. After this truncation, the above hierarchy of moment equations
becomes a closed set of coupled equations which can be recursively solved. Thus, given that the results derived in Ref.\ \cite{HTG17} are approximated, it is interesting to revisit the problem and get exact expressions of the rheological properties by considering both the Boltzmann equation for IMM and a BGK-type kinetic model for IHS. This will be carried out in the next two sections.

\section{Inelastic Maxwell models}
\label{sec3}

We consider in this section the Boltzmann equation \eqref{2.15} for IMM. In this case, the Boltzmann collision operator $J_\text{IMM}[f,f]$ is given by \cite{BK03}
\beq
\label{3.1}
J_\text{IMM}\left[{\bf v}_{1}|f,f\right] =\frac{\nu_\text{M}}{n \Omega_d} \int \; \dd{\bf v}_{2}\int
\dd\widehat{\boldsymbol{\sigma}} \left[ \alpha^{-1}f({\bf v}_{1}'')f({\bf v}_{2}'')-
f({\bf v}_{1})f({\bf v}_{2})\right],
\end{equation}
where $\Omega_d=2\pi^{d/2}/\Gamma(d/2)$ is the total solid angle in $d$ dimensions and $\nu_\text{M}$ is a collision frequency. In addition, the double primes on the velocities denote initial values $\left\{\mathbf{v}_1'',\mathbf{v}_2''\right\}$ that lead to $\left\{\mathbf{v}_1,\mathbf{v}_2\right\}$ following a binary collision:
\beq
\label{3.2}
\mathbf{v}_1''=\mathbf{v}_1-\frac{1}{2}\left(1+\al^{-1}\right)(\widehat{\boldsymbol{\sigma}}\cdot \mathbf{g})\widehat{\boldsymbol{\sigma}}, \quad \mathbf{v}_2''=\mathbf{v}_2+\frac{1}{2}\left(1+\al^{-1}\right)(\widehat{\boldsymbol{\sigma}}\cdot \mathbf{g})\widehat{\boldsymbol{\sigma}},
\eeq
where $\mathbf{g}=\mathbf{v}_1-\mathbf{v}_2$ is the relative velocity of the colliding pair and $\widehat{\boldsymbol{\sigma}}$ is a unit vector directed along the centers of the two colliding particles. The collision frequency $\nu_\text{M}(\mathbf{r},t)$ is independent of velocity but depends on space and time through its dependence on density and temperature. It can be seen as a free parameter of the model that can be chosen to optimize agreement with the properties of interest of the original Boltzmann equation for IHS. For instance, in order to correctly describe the velocity dependence of the original IHS collision rate, we can assume that the IMM collision rate is proportional to $T^{1/2}$.

As noted in previous works on IMM \cite{GS07,SG07}, the main advantage of the Boltzmann equation for Maxwell models (both elastic and inelastic) is that the moments of the operator $J[f,f]$ can be \emph{exactly} expressed in terms of the velocity moments of the velocity distribution $f$, without the knowledge of the latter. This property has been exploited to determine for arbitrary dimensions the explicit forms for all the second, third, and fourth-degree collisional moments as functions of the coefficient of restitution $\al$ \cite{GS07}. In the steady USF problem, the relevant velocity moments are the second- and fourth-degree moments since the third-degree moments vanish by symmetry. In particular, the second-degree collisonal moment (which is needed to get the rheological properties) is given by \cite{GS07}
\beq
\label{3.3}
\int\; \dd \mathbf{V} m V_iV_j J_\text{IMM}[\mathbf{V}|f,f]=-\nu_{0|2}\Pi_{ij}-p\zeta \delta_{ij},
\eeq
where $\Pi_{ij}=P_{ij}-p\delta_{ij}$ is the traceless part of the pressure tensor, $p=(P_{xx}+P_{yy}+\cdots)/d=nT$ is the hydrostatic pressure, and
\beq
\label{3.4.1}
\zeta=\frac{1-\al^2}{2d}\nu_\text{M},
\eeq
\beq
\label{3.4.2}
\nu_{0|2}=\zeta+\frac{(1+\al)^2}{2(d+2)}\nu_\text{M}=\frac{(d+1-\al)(1+\al)}{d(d+2)}\nu_\text{M}.
\eeq
The expressions of the fourth-degree collisional moments are displayed in the Appendix \ref{appA} for the sake of completeness. Equation \eqref{3.4.1} provides the exact form of the cooling rate for IMM. This form can be used to fix the value of the free parameter  $\nu_\text{M}$. This is chosen under the criterion that $\zeta$ of IMM is the same as that of IHS of diameter $\sigma$. Given that the cooling rate cannot be exactly evaluated for IHS, we take here for $\zeta_{\text{IHS}}$ its expression when $f$ is replaced by the Maxwellian distribution. In this approximation,  $\zeta_{\text{IHS}}$ is given by \cite{NE98}
\beq
\label{3.5}
\zeta_{\text{IHS}}\to \frac{d+2}{4d}(1-\al^2)\nu_0,
\eeq
where
\beq
\label{3.6}
\nu_0=\frac{8}{d+2}\frac{\pi^{(d-1)/2}}{\Gamma\left(\frac{d}{2}\right)}
n\sigma^{d-1}\sqrt{\frac{T}{m}}
\eeq
is the collision frequency of the shear viscosity coefficient of a dilute ordinary gas. Comparing Eqs.\ \eqref{3.4.1} and \eqref{3.5}, one gets the relationship
\beq
\label{3.7}
\nu_\text{M}=\frac{d+2}{2}\nu_0.
\eeq

\subsection{Rheological properties}

The hierarchy of equations defining the elements of the pressure tensor $P_{k\ell}$ can be easily obtained by multiplying both sides of Eq.\ \eqref{2.15} (replacing $J$ by $J_\text{IMM}$) by $m V_k V_\ell$ and integrating over $\mathbf{V}$. The result is
\beq
\label{3.8}
a \left(\delta_{kx}P_{\ell j}+\delta_{\ell x}P_{k y}\right)+2\gamma \left(P_{k\ell}-nT_\text{ex}\delta_{k\ell}\right)=-\nu_{0|2}P_{k\ell}-p\left(\zeta-\nu_{0|2}\right)\delta_{k\ell},
\eeq
where use has been made of Eq.\ \eqref{3.3}. From Eq.\ \eqref{3.8} is easy to prove that the diagonal elements of the pressure tensor orthogonal to the shear plane $xy$ are equal to $P_{yy}$ (i.e., $P_{yy}=P_{zz}= \ldots=P_{dd}$). As a consequence, $P_{xx}=d p-(d-1)P_{yy}$ and the elements $P_{yy}$ and $P_{xy}$ obey the equations
\beq
\label{3.9}
\left(\nu_{0|2}+2\gamma\right)P_{yy}=n T_\text{ex}\left[2\gamma-\left(\zeta-\nu_{0|2}\right)\theta\right],
\eeq
\beq
\label{3.10}
\left(\nu_{0|2}+2\gamma\right)P_{xy}=-aP_{yy},
\eeq
where we recall that $\theta\equiv T/T_\text{ex}$. The solution to Eqs.\ \eqref{3.9} and \eqref{3.10} is
\beq
\label{3.11}
P_{yy}=\frac{2\gamma-\left(\zeta-\nu_{0|2}\right)\theta}{\nu_{0|2}+2\gamma}nT_{\text{ex}},
\eeq
\beq
\label{3.12}
P_{xy}=-\frac{a}{\nu_{0|2}+2\gamma}P_{yy}=-\frac{2\gamma-\left(\zeta-\nu_{0|2}\right)\theta}{\left(\nu_{0|2}+2\gamma\right)^2}anT_{\text{ex}}.
\eeq
The element $P_{xx}$ can be easily obtained from Eq.\ \eqref{3.11} as
\beq
\label{3.13}
P_{xx}=\frac{d\left(\nu_{0|2}+2\gamma\right)\theta-(d-1)\left[2\gamma-\left(\zeta-\nu_{0|2}\right)\theta\right]}{\nu_{0|2}+2\gamma}nT_{\text{ex}}.
\eeq
The (reduced) temperature $\theta$ can be finally determined by substituting Eq.\ \eqref{3.12} into the steady-state condition \eqref{2.16}. In order to compare our theoretical results with those obtained in Ref.\ \cite{HTG17} by computer simulations, it is convenient to scale the shear rate with the friction coefficient $\gamma$ (i.e., $a^*\equiv a/\gamma$) and introduce the (reduced) background gas temperature $T_\text{ex}^*\equiv T_\text{ex}/m\sigma^2 \gamma^2$. In terms of these quantities, the solution to Eq.\ \eqref{2.16} can be written as
\beq
\label{3.14}
a^*=\sqrt{\frac{d}{2}\frac{\sqrt{\theta}\zeta^*+2(1-\theta^{-1})}{\sqrt{\theta}(\nu_{0|2}^*-\zeta^*)+2\theta^{-1}}}(2+\sqrt{\theta}\nu_{0|2}^*),
\eeq
where we have introduced the dimensionless quantities
\beq
\label{3.15}
\zeta^*\equiv \frac{\zeta}{\sqrt{\theta}\gamma}=\frac{2\pi^{(d-1)/2}}{d\Gamma\left(\frac{d}{2}\right)} (1-\al^2)n^* \sqrt{T_{\text{ex}}^*},
\eeq
\beq
\label{3.16}
\nu_{0|2}^*\equiv \frac{\nu_{0|2}}{\sqrt{\theta}\gamma}=\frac{4\pi^{(d-1)/2}}{d(d+2)\Gamma\left(\frac{d}{2}\right)}(d+1-\al)(1+\al) n^* \sqrt{T_{\text{ex}}^*}.
\eeq
Since $\gamma\propto \sqrt{T_{\text{ex}}}$, then $\zeta^*$ and $\nu_{0|2}^*$ are independent of both the granular temperature $T$ and the background temperature $T_\text{ex}$. In Eqs.\ \eqref{3.15} and \eqref{3.16}, $n^*=n\sigma^d$ is the reduced density. Note that this explicit dependence on density comes from the scaling of the shear rate $a$ and the bath temperature $T_\text{ex}$. If we had reduced the shear rate for instance with the collision frequency $\nu_0(T)$, then the above density dependence had been removed. On the other hand, since we want to make a close comparison with the simulation data reported in Ref.\ \cite{HTG17}, our theory must employ the same input parameters as in the simulation results.

As happens for IHS \cite{HTG17}, it is quite apparent that we cannot express the (reduced) temperature $\theta$ in Eq.\ \eqref{3.14} as an explicit function of both the coefficient of restitution $\al$ and the (reduced) shear rate $a^*$. However, the dependence of $\theta$ on the latter parameters can implicitly be obtained from the physical solution to Eq.\ \eqref{3.14} as $a^{*2}(\theta,\al)$. Once $\theta$ is known, the remaining rheological functions can be determined from Eqs.\ \eqref{3.11} and \eqref{3.12} in terms of $\al$ and $a^*$. In particular, the (dimensionless) non-Newtonian shear viscosity
\beq
\label{3.16.1}
\eta^*\equiv -\frac{P_{xy}^*}{a^*}
\eeq
can be easily identified from Eq.\ \eqref{3.12} with the result
\beq
\label{3.17}
\eta^*=\frac{2+\left(\nu_{0|2}^*-\zeta^*\right)\sqrt{\theta}\theta}{\left(\sqrt{\theta}\nu_{0|2}^*+2\right)^2}.
\eeq
Since $P_{yy}=P_{zz}$, the only nonzero (reduced) viscometric function is given by
\beq
\label{3.18}
\Psi^*=\frac{P_{xx}-P_{yy}}{nT_\text{ex}}=d\theta \frac{2\left(1-\theta^{-1}\right)+\sqrt{\theta}\zeta^*}{2+\sqrt{\theta}\nu_{0|2}^*},
\eeq
where use has been made of Eqs.\ \eqref{3.11} and \eqref{3.13}. It must be remarked that although the theoretical prediction $P_{yy}=P_{zz}$ disagrees with computer simulations \cite{HTG17}, the magnitude of the difference $P_{yy}-P_{zz}$ is in general very small; therefore the expressions \eqref{3.11}--\eqref{3.13} can be still considered as reliable. A careful comparison with the theoretical results obtained for dense granular suspensions of IHS \cite{HTG17} by means of Grad's moment method shows that these expressions differ from those derived here for IMM in the dilute limit. On the other hand, this discrepancy is only due to the different $\al$ dependence of the eigenvalue $\nu_{0|2}$ with respect to the one found for IHS.

For illustrative purposes, it is interesting to consider the limits of small and large shear rates. For small shear rates ($a^*\to 0$), $\eta^*\to \eta_\text{NS}^*$, where the Navier--Stokes shear viscosity of the granular suspension is
\beq
\label{3.19}
\eta_\text{NS}^*=\frac{\theta_\text{NS}}{2+\sqrt{\theta_\text{NS}}\nu_{0|2}^*}.
\eeq
Here, $\theta_\text{NS}$ is a real solution of the equation
\beq
\label{3.20}
\theta_\text{NS}=\frac{1}{1+\frac{1}{2}\sqrt{\theta_\text{NS}}\zeta^*}.
\eeq
For large shear rates ($a^*\to \infty$), the asymptotic forms for $\al<1$ are
\beq
\label{3.21}
\theta_{\infty}\to \frac{2}{d}\frac{\nu_{0|2}^*-\zeta^*}{\nu_{0|2}^{*2}\zeta^*}a^{*2}, \quad
\eta_{\infty}^*\to \sqrt{\frac{d}{2}} \frac{\left(\nu_{0|2}^*-\zeta^*\right)^{3/2}}{\nu_{0|2}^{*3}\sqrt{\zeta^*}}a^{*},
\eeq
while for elastic collisions ($\al=1$), one gets
\beq
\label{3.22}
\theta_{\infty}\to \frac{a^{*4}}{d^2\nu_{0|2}^{*2}}, \quad
\eta_{\infty}^*\to \frac{a^{*2}}{d\nu_{0|2}^{*2}}.
\eeq

\begin{figure}
{\includegraphics[width=0.4\columnwidth]{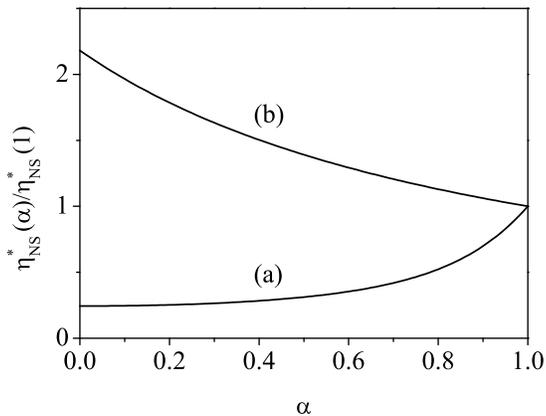}}
\caption{Plot of the ratios $\eta_\text{NS}^*(\al)/\eta_\text{NS}^*(1)$ (a) and $\eta^*_\text{NS,dry}(\al)/\eta^*_\text{NS,dry}(1)$
(b) as functions of the coefficient of restitution $\al$ for a three-dimensional system.
\label{fig1}}
\end{figure}

\vicente{It is interesting at this point to compare the behaviors of the non-Newtonian shear viscosity obtained here for granular suspensions in the limit of small and high shear rates with those derived before for \emph{dilute} ordinary \cite{GS03} and dry inelastic \cite{SG07} Maxwell gases. In both cases, while $\eta^*\equiv \text{finite}$ when $a^*\to 0$, $\eta^* \propto a^{*-4/3}$ when $a^*\to \infty$. This means that $\eta^*$ is a monotonically decreasing function of the shear rate and so, the shearing produces an inhibition of the momentum transport (shear thinning effect) in the sense that the actual value of the shear stress $|P_{xy}|$ is smaller than the one predicted by Newton's law. On the other hand, a completely different behavior is found here for granular suspensions, since while $\eta^*\equiv \text{finite}$ when $a^*\to 0$, this coefficient diverges in the limit $a^*\to \infty$ [see Eqs.\ \eqref{3.21} and \eqref{3.22}]. Thus, the fact that the ratio $\eta^*(a^*\to \infty)/\eta^*(a^*\to 0)$ becomes very large could explain the existence of discontinuous shear thickening in a structurally simple system due to the connection between the Newtonian and Bagnoldian branches. This behavior changes as the density of the system increases since kinetic theory results predict \emph{continuous} shear thickening for both ordinary gases \cite{SMDB98} and granular suspensions \cite{HTG17}.}

Before considering the shear-rate dependence of the rheological functions, it is worthwhile to compare the $\al$-dependence of the Navier--Stokes shear viscosity \eqref{3.19} with the one obtained in the \emph{dry} granular case. In dimensionless form, the expression of the Navier--Stokes shear viscosity of a granular gas can be written as $\eta_\text{NS,dry}=(p/\nu_\text{M})\eta^*_\text{NS,dry}$, where \cite{S03}
\beq
\label{3.23}
\eta^*_\text{NS,dry}=\frac{4d(d+2)}{(1+\al)\left[3d+2-(d-2)\al\right]}.
\eeq
Figure \ref{fig1} shows the ratios $\eta_\text{NS}^*(\al)/\eta_\text{NS}^*(1)$ and $\eta^*_\text{NS,dry}(\al)/\eta^*_\text{NS,dry}(1)$ as functions of the coefficient of restitution $\al$ for $d=3$. Here, $\eta_\text{NS}^*(1)$ and $\eta^*_\text{NS,dry}(1)$ refer to the values of the shear viscosity coefficients for elastic collisions for the suspension and dry granular cases, respectively. It is quite apparent that the $\al$ dependence of both viscosities is qualitatively different since while the shear viscosity of a granular suspension decreases (with respect to its value for elastic collisions) with increasing inelasticity, the opposite happens for granular gases. Moreover, the impact of inelasticity on both shear viscosity coefficients is quite significant.

\subsection{Fourth-degree moments}

As mentioned in section \ref{sec1}, although the rheological properties are the most important transport properties of the granular suspension, the determination of higher degree velocity moments is also an appealing problem. Since the third-degree moments vanish in the steady USF by symmetry reasons, the fourth-degree moments are the first nonzero moments beyond the second-degree moments. Here, we will focus on a three-dimensional system ($d=3$). As for ordinary gases \cite{SG95,GS03}, for $d=3$, there are 15 independent fourth-degree moments; 6 are \emph{asymmetric} (in the sense that they vanish in the steady state) and 9 are \emph{symmetric} (they are different from zero in the steady state). The symmetric and asymmetric moments are uncoupled. Since we are not interested in this paper in analyzing the time evolution of the fourth-degree moments, we will address here only the study of the (steady) symmetric moments.

In parallel to the elastic case \cite{SG95,GS03}, we choose the following set of 9 symmetric moments:
\beq
\label{3.24}
\Big\{M_{4|0}, M_{2|xx}, M_{2|yy}, M_{2|xy}, M_{0|xxxx}, M_{0|yyyy}, M_{0|zzzz}, M_{0|xxxy}, M_{0|xyyy}\Big\}.
\eeq
Here, we have introduced the velocity moments
\beq
\label{a4}
\Big(M_{4|0}, M_{2|ij}, M_{0|ijk\ell}\Big)=\int \dd\mathbf{V}\; \Big(Y_{4|0}, Y_{2|ij}, Y_{0|ijk\ell}\Big) f(\mathbf{V}),
\eeq
where the fourth-degree Ikenberry polynomials are defined as \cite{TM80}
\beq
\label{a1}
Y_{4|0}(\mathbf{V})=V^4, \quad Y_{2|ij}(\mathbf{V})=V^2\left(V_i V_j-\frac{1}{3}V^2\delta_{ij}\right),
\eeq
\beqa
\label{a2}
Y_{0|ijk\ell}(\mathbf{V})&=&V_iV_jV_kV_\ell-\frac{V^2}{7}\bigg(V_iV_j\delta_{k\ell}+V_iV_k\delta_{j\ell}+V_iV_\ell\delta_{jk}
+V_jV_jk\delta_{i\ell}+V_jV_\ell\delta_{ik}
\nonumber\\
& & +V_kV_\ell\delta_{ij}\bigg)+\frac{V^4}{35}\bigg(\delta_{ij}\delta_{k\ell}+\delta_{ik}\delta_{j\ell}+\delta_{i\ell}\delta_{jk}
\bigg).
\eeqa
As for ordinary gases \cite{SG95,GS03}, it is easy to prove that the combination
\beq
\label{3.25}
3M_{0|xxxx}-4\left(M_{0|yyyy}+M_{0|zzzz}\right)=0
\eeq
in the steady USF state. This means that we really have 8 independent fourth-degree symmetric moments since for instance $M_{0|xxxx}=\frac{4}{3}\left(M_{0|yyyy}+M_{0|zzzz}\right)$. As expected, the eight independent moments are coupled. The corresponding equations obeying those eight moments can be determined by multiplying both sides of Eq.\ \eqref{2.15} by the set of velocity polynomials
\beq
\label{3.26}
\Big\{Y_{4|0}, Y_{2|xx}, Y_{2|yy}, Y_{2|xy}, Y_{0|yyyy}, Y_{0|zzzz}, Y_{0|xxxy}, Y_{0|xyyy}\Big\}
\eeq
and integrating over velocity. In addition, to explicitly obtain the hierarchy of moment equations, one needs the collisional moments \eqref{a6}--\eqref{a10} associated with the above fourth-degree polynomials. In dimensionless form, the set of coupled equations for the fourth-degree moments can be written in matrix form as
\beq
\label{3.27}
\mathcal{L}_{\mu\nu}\mathcal{M}_{\nu}=\mathcal{N}_{\mu}, \quad \mu=1,2,\ldots,8.
\eeq
Here, $\boldsymbol{\mathcal{M}}$ is the column matrix defined by the set
\beq
\label{3.28}
\Big\{M_{4|0}^*, M_{2|xx}^*, M_{2|yy}^*, M_{0|yyyy}^*, M_{0|zzzz}^*, M_{2|xy}^*, M_{0|xxxy}^*, M_{0|xyyy}^*\Big\},
\eeq
and $\boldsymbol{\mathcal{L}}$ is the square matrix
\beq
\label{3.29}
\boldsymbol{\mathcal{L}}=4\boldsymbol{I}+\boldsymbol{\mathcal{L}}',
\eeq
where $\boldsymbol{I}$ is the $8 \times 8$ unit matrix and
\begin{equation}
\label{3.30}
\boldsymbol{\mathcal{L}}'=\left(\begin{array}{cccccccc}
	\sqrt{\theta}\nu^*_{4|0} & 0 & 0 & 0 & 0 & 4a^* & 0 & 0 \\
	0 & \sqrt{\theta}\nu^*_{2|2}& 0 & 0 & 0 & \frac{32}{21}a^* & 2a^* & 0 \\
	0 & 0 & \sqrt{\theta}\nu^*_{2|2} & 0 & 0 & -\frac{10}{21}a^* & 0 & 2a^* \\
	0 & 0 & 0 & \sqrt{\theta}\nu^*_{0|4} & 0 & -\frac{96}{245}a^* & 0 & -\frac{12}{7}a^* \\
	0 & 0 & 0 & 0 & \sqrt{\theta}\nu^*_{0|4}& \frac{24}{245}a^* & \frac{12}{7}a^* & \frac{12}{7}a^* \\
	\frac{7}{15}a^* & \frac{2}{7}a^* & \frac{9}{7}a^* & -\frac{7}{3}a^* & -\frac{1}{3}a^* &\sqrt{\theta}\nu^*_{2|2}& 0 & 0 \\
	0 & \frac{15}{49}a^* & -\frac{6}{49}a^* & -\frac{5}{2}a^* & -\frac{5}{14}a^* & 0 &\sqrt{\theta}\nu^*_{0|4}& 0 \\
	0 & -\frac{6}{49}a^* & \frac{15}{49}a^* & 2a^* & \frac{1}{7}a^* & 0 & 0 &\sqrt{\theta}\nu^*_{0|4}
\end{array}\right).
\end{equation}
The scaled moments $M_{4|0}^*$, $M_{2|ij}^*$, and $M_{0|ijk\ell}^*$ are defined as
\beq
\label{3.30.1}
\left\{M_{4|0}^*, M_{2|ij}^*, M_{0|ijk\ell}^*\right\} = n^{-1}\left(\frac{m}{T_\text{ex}}\right)^2 \left\{M_{4|0}, M_{2|ij}, M_{0|ijk\ell}\right\},
\eeq
and in Eq.\ \eqref{3.30}, $\nu_{4|0}^*\equiv\nu_{4|0}/(\sqrt{\theta}\gamma)$, $\nu_{2|2}^*\equiv\nu_{2|2}/(\sqrt{\theta}\gamma)$, and $\nu_{0|4}^*\equiv\nu_{0|4}/(\sqrt{\theta}\gamma)$. The expressions of $\nu_{4|0}$, $\nu_{2|2}$, and $\nu_{0|4}$ are given by Eqs.\ \eqref{a11} and \eqref{a12}, respectively. In addition, the elements of the column matrix $\boldsymbol{\mathcal{N}}$ are made of second-degree moments:
\beq
\label{3.31}
\mathcal{N}_1=9\theta^2\sqrt{\theta}\lambda_1^*-2\sqrt{\theta}\lambda_2^*\left(3\Pi_{yy}^{*2}+\Pi_{xy}^{*2}\right)+60\theta,
\eeq
\beq
\label{3.32}
\mathcal{N}_2=-6\theta\sqrt{\theta}\lambda_3^*\Pi_{yy}^*-\frac{\sqrt{\theta}}{3}\lambda_4^*\left(2\Pi_{yy}^{*2}-\Pi_{xy}^{*2}\right)-28\Pi_{yy}^*,
\eeq
\beq
\label{3.33}
\mathcal{N}_3=3\theta\sqrt{\theta}\lambda_3^*\Pi_{yy}^*+\frac{\sqrt{\theta}}{3}\lambda_4^*\left(3\Pi_{yy}^{*2}+\Pi_{xy}^{*2}\right)+14\Pi_{yy}^*,
\eeq
\beq
\label{3.34}
\mathcal{N}_4=\frac{3}{35}\sqrt{\theta}\lambda_5^*\left(27\Pi_{yy}^{*2}-16\Pi_{xy}^{*2}\right),
\eeq
\beq
\label{3.35}
\mathcal{N}_5=\frac{3}{35}\sqrt{\theta}\lambda_5^*\left(27\Pi_{yy}^{*2}+4\Pi_{xy}^{*2}\right),
\eeq
\beq
\label{3.36}
\mathcal{N}_6=3\theta\sqrt{\theta}\lambda_3^*\Pi_{xy}^*+\sqrt{\theta}\lambda_4^*\Pi_{yy}^{*}\Pi_{xy}^{*}+14\Pi_{xy}^*,
\eeq
\beq
\label{3.37}
\mathcal{N}_7=-\frac{36}{7}\sqrt{\theta}\lambda_5^*\Pi_{yy}^{*}\Pi_{xy}^{*},
\eeq
\beq
\label{3.38}
\mathcal{N}_8=\frac{27}{7}\sqrt{\theta}\lambda_5^*\Pi_{yy}^{*}\Pi_{xy}^{*},
\eeq
where $\lambda_i^*\equiv \lambda_i/(\sqrt{\theta}\gamma)$ and $\Pi_{ij}^*\equiv \Pi_{ij}/nT_\text{ex}$. The quantities $\lambda_i (i=1,\cdots,5)$ are defined by Eqs.\ \eqref{a13} and \eqref{a14}.

The solution to Eq.\ \eqref{3.27} is
\beq
\label{3.39}
\boldsymbol{\mathcal{M}}=\boldsymbol{\mathcal{L}}^{-1}\cdot \boldsymbol{\mathcal{N}}.
\eeq
Equation \eqref{3.39} provides the dependence of the (symmetric) fourth-degree moments on both the (reduced) shear rate $a^*$ and the coefficient of restitution $\al$. This dependence will be analyzed in section \ref{sec5}.

\section{BGK-type kinetic model of the Boltzmann equation}
\label{sec4}

We consider now the results derived for the USF from a BGK-type kinetic model of the Boltzmann equation \cite{BDS99}. In the USF problem, the steady kinetic model for the granular suspension described by the Boltzmann equation \eqref{2.15} reads
\beq
\label{4.1}
-aV_y\frac{\partial f}{\partial V_x}-\gamma \frac{\partial}{\partial
{\bf V}}\cdot {\bf V} f-\frac{\gamma T_\text{ex}}{m}\frac{\partial^2 f}{\partial V^2} =-\chi(\alpha)\nu_0 \left(f-f_\text{L}\right)+\frac{\zeta}{2}\frac{\partial}{\partial
{\bf V}}\cdot {\bf V} f,
\eeq
where $\nu_0$ is the effective collision frequency defined by Eq.\ \eqref{3.6}, $\zeta$ is defined by Eq.\ \eqref{3.4.1} [or equivalently, by Eq.\ \eqref{3.5}] and
\begin{equation}
\label{4.2}
f_\text{L}(\mathbf{V})=n\left(\frac{m}{2\pi T}\right)^{d/2}e^{-mV^2/2T}
\end{equation}
is the local equilibrium distribution function In addition, $\chi(\alpha)$ is a free parameter of the model chosen to optimize the agreement with the Boltzmann results.

One of the main advantages of using a kinetic model instead of the Boltzmann equation is that it lends itself to determine all the velocity moments of the velocity distribution function. For the sake of convenience, let us define the general velocity moments
\beq
\label{4.3}
M_{k_1,k_2,k_3}=\int\; \dd \mathbf{V}\;  V_x^{k_1}V_y^{k_2}V_z^{k_3} f(\mathbf{V}).
\eeq
As for IMM, although we are mainly interested in  the three-dimensional case, we will perform our results for $d=3$ and $d=2$.  Of course, for hard disks ($d=2$), $k_3=0$ since the $z$-axis is meaningless. To get $M_{k_1,k_2,k_3}$, we multiply both sides of Eq.\ \eqref{4.1} by $V_x^{k_1}V_y^{k_2}V_z^{k_3}$ and integrate over velocity to achieve the result
\beq
\label{4.4}
a k_1 M_{k_1-1,k_2+1,k_3}+\left(\chi \nu_0+k \lambda\right)M_{k_1,k_2,k_3}=N_{k_1,k_2,k_3},
\eeq
where $\lambda=\gamma+\zeta/2$, $k=k_1+k_2+k_3$, and
\beq
\label{4.5}
N_{k_1,k_2,k_3}=\frac{\gamma T_\text{ex}}{m} R_{k_1,k_2,k_3}+\chi \nu_0 M_{k_1,k_2,k_3}^\text{L}.
\eeq
In Eq.\ \eqref{4.5}, we have introduced the quantities
\beqa
\label{4.6}
R_{k_1,k_2,k_3}&=&\int\; \dd \mathbf{V}\; f(\mathbf{V})\frac{\partial^2}{\partial V^2}\left(V_x^{k_1}V_y^{k_2}V_z^{k_3}\right)\nonumber\\
&=&
k_1(k_1-1)M_{k_1-2,k_2,k_3}+k_2(k_2-1)M_{k_1,k_2-2,k_3}+k_3(k_3-1)M_{k_1,k_2,k_3-2},
\eeqa
and
\beq
\label{4.7}
M_{k_1,k_2,k_3}^\text{L}=n \left(\frac{2T}{m}\right)^{k/2}\pi^{-d/2}\Gamma\left(\frac{k_1+1}{2}\right)\Gamma\left(\frac{k_2+1}{2}\right)
\Gamma\left(\frac{k_3+1}{2}\right)
\eeq
if $k_1$, $k_2$, and $k_3$ are even, being zero otherwise. The solution to Eq.\ \eqref{4.4} can be cast into the form (see the Appendix \ref{appB})
\beq
\label{4.8}
M_{k_1,k_2,k_3}=\sum_{q=0}^{k_1} \frac{k_1!}{(k_1-q)!} \frac{(-a)^q}{\left(\chi \nu_0+k \lambda\right)^{1+q}}N_{k_1-q,k_2+q,k_3}.
\eeq

The first nontrivial moments are related with the pressure tensor $P_{ij}$. The expressions of its nonzero elements are
\beq
\label{4.9}
P_{yy}=P_{zz}=n T_\text{ex}\frac{\theta \chi \nu_0+2\gamma}{\chi \nu_0+2\lambda}, \quad P_{xy}=-n T_\text{ex}\frac{\theta \chi \nu_0+2\gamma}{\left(\chi \nu_0+2\lambda\right)^2}a,
\eeq
\beq
\label{4.10}
P_{xx}= d n T-(d-1)P_{yy}=n T_\text{ex}\frac{\theta \chi \nu_0+2\gamma}{\chi \nu_0+2\lambda}\left[1+\frac{2a^2}{\left(\chi \nu_0+2\lambda\right)^2}\right].
\eeq
The non-Newtonian shear viscosity $\eta^*$ and the viscometric function $\Psi^*$ defined by Eqs.\ \eqref{3.16.1} and \eqref{3.17}, respectively, can be easily identified from Eqs.\ \eqref{4.9} and \eqref{4.10}. Their expressions in the BGK model are
\beq
\label{4.11}
\eta^*=\frac{2+\chi \nu_0^*\sqrt{\theta}\theta}{\left[\sqrt{\theta}\left(\chi \nu_0^*+\zeta^*\right)+2\right]^2},
\eeq
\beq
\label{4.12}
\Psi^*=d\theta \frac{2\left(1-\theta^{-1}\right)+\sqrt{\theta}\zeta^*}{2+\sqrt{\theta}\left(\chi \nu_0^*+\zeta^*\right)},
\eeq
where $\zeta^*$ is given by Eq.\ \eqref{3.15} and
\beq
\label{4.13}
\nu_0^*\equiv \frac{\nu_0}{\sqrt{\theta}\gamma}=\frac{8}{d+2}\frac{\pi^{(d-1)/2}}{\Gamma\left(\frac{d}{2}\right)}n^* \sqrt{T_{\text{ex}}^*}.
\eeq
Finally, the steady granular temperature $\theta\equiv T/T_\text{ex}$ can be obtained from the steady-state condition \eqref{2.16}. After some algebra, one gets the implicit equation
\beq
\label{4.14}
a^*=\sqrt{\frac{d}{2}\frac{\sqrt{\theta}\zeta^*+2(1-\theta^{-1})}{\sqrt{\theta}\chi \nu_0^*+2\theta^{-1}}}\left[2+\sqrt{\theta}\left(\chi \nu_0^*+\zeta^*\right)\right].
\eeq
Comparison of Eqs.\ \eqref{4.11}, \eqref{4.12}, and \eqref{4.14} with those obtained by solving the Boltzmann equation via Grad's moment method \cite{HTG17} shows that the BGK results for the rheological properties agree with the Boltzmann ones when the parameter $\chi(\al)$ is given by
\beq
\label{4.15}
\chi(\al)=\frac{1+\al}{2}\left[1-\frac{d-1}{2d}(1-\al)\right].
\eeq
Furthermore, for elastic collisions, Eqs.\ \eqref{4.11}, \eqref{4.12}, and \eqref{4.14} agree with previous results \cite{HT17} obtained by solving the BGK model for ordinary dilute gases by means of Grad's moment method.

The expressions of the fourth-degree moments can be easily obtained from Eq.\ \eqref{4.8} with the choice \eqref{4.15}. The shear-rate dependence of these moments will be compared with the ones derived before for IMM for $d=3$ in section \ref{sec5}.

\subsection{Transport properties at $T_\text{ex}=0$}

Apart from getting the velocity moments, the use of the BGK equation allow us in some cases to obtain explicitly the velocity distribution function $f$. On the other hand, we have not been able to derive an expression for $f$ for the suspension model \eqref{4.1}. An exception corresponds to the simple limit case $T_\text{ex}=0$ but keeping $\gamma\equiv \text{const}$. It corresponds to a situation where the background temperature $T_\text{ex}$ is much smaller than the granular temperature $T$ and hence, the model ignores the effects of thermal fluctuations on solid particles and the impact of the gas phase is only accounted for by the drag force term. Of course, it is also understood that $\gamma$ does not depend on the background temperature. This simple model has been employed in several previous works to study simple shear flows in gas-solid flows  \cite{TK95,SMTK96,SA17,ChVG15}, particle clustering due to hydrodynamic interactions \cite{WK00}, steady states of particle systems driven by a vibrating boundary \cite{WZLH09} and more recently \cite{H13,HT13,SMMD13,WGZS14} to analyze the rheology of frictional sheared hard-sphere suspensions.

\vicente{Note that, in spite of the absence of the Langevin-like term $T_\text{ex} \partial^2 f/\partial v^2$ in this suspension model, the Boltzmann equation \eqref{2.4} still admits a simple solution in the homogeneous state (zero shear rate) for elastic collisions ($\al=1$). Thus, if one chooses a convenient selection of frame then $\mathbf{U}=\mathbf{U}_g=\mathbf{0}$, and Eq.\ \eqref{2.4} admits the time-dependent solution
\beq
\label{4.15.0}
f_\text{L}(\mathbf{v},t)=n\left(\frac{m}{2\pi T(t)}\right)^{d/2}e^{-mv^2/2T(t)},  
\eeq
where $T(t)$ verifies the equation
\beq
\label{4.15.1}
\frac{\partial \ln T}{\partial t}=-2\gamma.
\eeq
An $H$-theorem has been also proved \cite{GSB90} for this time-dependent Maxwellian distribution in the sense that, starting
from any initial condition and in the presence of the viscous drag force $\gamma \mathbf{v}$, the velocity distribution function $f (\mathbf{r},\mathbf{v},t)$ reaches in the long time limit the Maxwellian form with a time-dependent temperature.}

In this limit case ($T_\text{ex}=0$), according to Eqs.\ \eqref{4.9} and \eqref{4.10}, the elements of the pressure tensor can be written in a more compact form as \cite{G17}
\beq
\label{4.16}
P_{yy}=P_{zz}=\frac{n T}{1+2\xi}, \quad P_{xx}=d n T-(d-1)P_{yy}, \quad P_{xy}=-\frac{n T}{(1+2\xi)^2}\widetilde{a},
\eeq
where $\widetilde{a}=a/(\nu_0\chi)$, and $\xi$ is the real root of the cubic equation
\beq
\label{4.17}
d\xi(1+2\xi)^2=\widetilde{a}^2,
\eeq
namely,
\beq
\label{4.18}
\xi(\widetilde{a})=\frac{2}{3}\sinh^2\left[\frac{1}{6}\cosh^{-1}\left(1+\frac{27}{d}\widetilde{a}^2\right)\right].
\eeq
The friction coefficient $\gamma$ obeys the steady-state condition \eqref{2.16}:
\beq
\label{4.19}
\gamma=\chi \nu_0 \xi-\frac{1}{2}\zeta.
\eeq
Since $\gamma \geq 0$, at a given value of $\al$, there is a critical value $\widetilde{a}_c(\al)$ of the (reduced) shear rate such that physical solutions to Eq.\ \eqref{4.15} only exist for $\widetilde{a}\geq \widetilde{a}_c(\al)$. The critical value $\widetilde{a}_c$ is obtained from the condition $2\chi \nu_0 \xi=\zeta$. Thus, if $\al \neq 1$, then $\widetilde{a}_c >0$ and the expression for the Newtonian shear viscosity cannot be recovered when $\widetilde{a} \to 0$. This is a drawback of this suspension model ($T_\text{ex}=0$). It must be remarked that Tsao and Koch \cite{TK95} solved time ago this simple model and showed the existence of a discontinuous transition for the temperature between a ``quenched'' state (a low temperature state) and an ``ignited'' state (a high temperature state).

Finally, the velocity distribution function $f(\mathbf{V})$ can be also determined explicitly in this limit case. When $T_\text{ex}=0$, the BGK equation \eqref{4.1} becomes
\beq
\label{4.20}
-aV_y\frac{\partial f}{\partial V_x}-\lambda \frac{\partial}{\partial
{\bf V}}\cdot {\bf V} f+\chi \nu_0 f = \chi \nu_0 f_\text{L}.
\eeq
This equation can be rewritten as
\beq
\label{4.21}
\left(1-d \widetilde{\lambda}-\widetilde{a}V_y\frac{\partial}{\partial V_x}-\widetilde{\lambda}\mathbf{V}\cdot \frac{\partial}{\partial
{\bf V}}\right)f=f_\text{L},
\eeq
where $\widetilde{\lambda}=\lambda/(\chi \nu_0)$. The hydrodynamic solution to Eq.\ \eqref{4.21} is
\beqa
\label{4.22}
f&=&\left(1-d \widetilde{\lambda}-\widetilde{a}V_y\frac{\partial}{\partial V_x}-\widetilde{\lambda}\mathbf{V}\cdot \frac{\partial}{\partial
{\bf V}}\right)^{-1} f_\text{L} \nonumber\\
&=&\int_0^{\infty}\; \dd s\; e^{-(1-d\widetilde{\lambda})s}\;
e^{\widetilde{a}sV_y \frac{\partial}{\partial V_x}}\; e^{\widetilde{\lambda}s\mathbf{V}\cdot \frac{\partial}{\partial \mathbf{V}}}
f_\text{L}(\mathbf{V}).
\eeqa
The action of the velocity operators $e^{\widetilde{a}sV_y \frac{\partial}{\partial V_x}}$ and
$e^{\widetilde{\lambda}t\mathbf{V}\cdot \frac{\partial}{\partial \mathbf{V}}}$ on an arbitrary function $g(\mathbf{V})$ is
\beq
\label{4.23}
e^{\widetilde{a}sV_y \frac{\partial}{\partial V_x}}g(V_x,V_y,V_z)=g(V_x+\widetilde{a}sV_y, V_y,V_z),
\eeq
\beq
\label{4.24}
e^{\widetilde{\lambda}s\mathbf{V}\cdot \frac{\partial}{\partial \mathbf{V}}}g(V_x,V_y,V_z)=
g\left(e^{\widetilde{\lambda}s}V_x,e^{\widetilde{\lambda}s}V_y,e^{\widetilde{\lambda}s}V_z\right).
\eeq
Taking into account these operators, the velocity distribution function $f$ can be finally written as
\beq
\label{4.25}
f(\mathbf{V})=n \left(\frac{m}{2T}\right)^{d/2}\varphi(\mathbf{c}),
\eeq
where $\mathbf{c}=(m/2T)^{1/2}\mathbf{V}$ is the reduced peculiar velocity and the reduced velocity distribution function $\varphi(\mathbf{c})$ is
\beqa
\label{4.26}
\varphi(\mathbf{c})&=&\pi^{-d/2}\int_0^{\infty}\; \dd s\;  e^{-(1-d\widetilde{\lambda})s}\;
\exp \left[-e^{2\widetilde{\lambda}s}\left(\mathbf{c}+s \; \widetilde{\mathbf{a}}\cdot \mathbf{c}\right)^2\right]\nonumber\\
&=&
\pi^{-d/2}\int_0^{\infty}\; \dd s\;  e^{-(1-d\widetilde{\lambda})s}\;
\exp\Big\{-e^{2\widetilde{\lambda}s}\big[(c_x+\widetilde{a}s c_y)^2+c_y^2+c_z^2)\big]\Big\}.
\eeqa
Here, we have introduced the tensor $\widetilde{a}_{ij}=\widetilde{a}\delta_{ix}\delta_{jy}$.

\section{Rheological properties and fourth-degree moments. Comparison with computer simulations}
\label{sec5}

\begin{figure}
{\includegraphics[width=0.4\columnwidth]{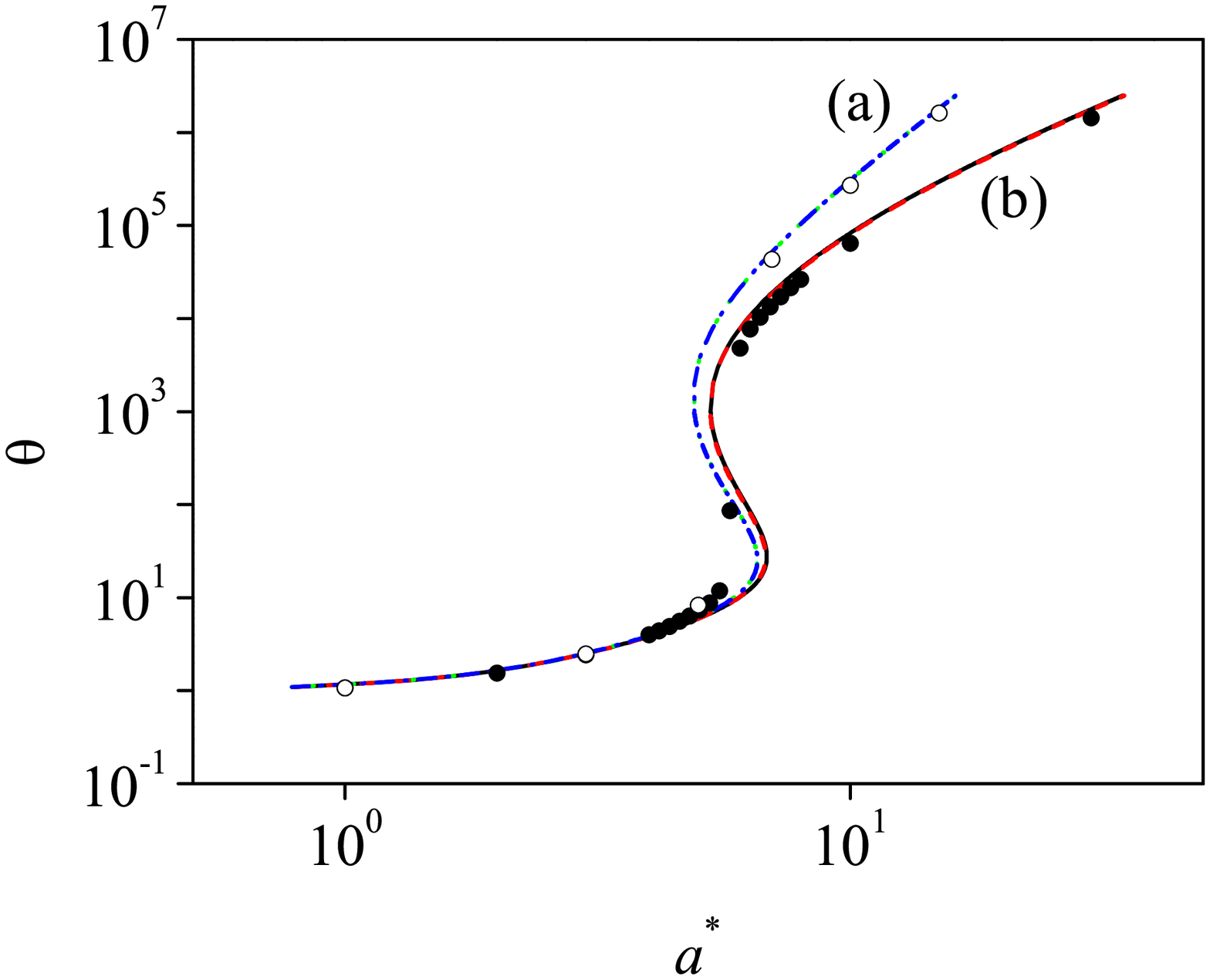}}
{\includegraphics[width=0.4\columnwidth]{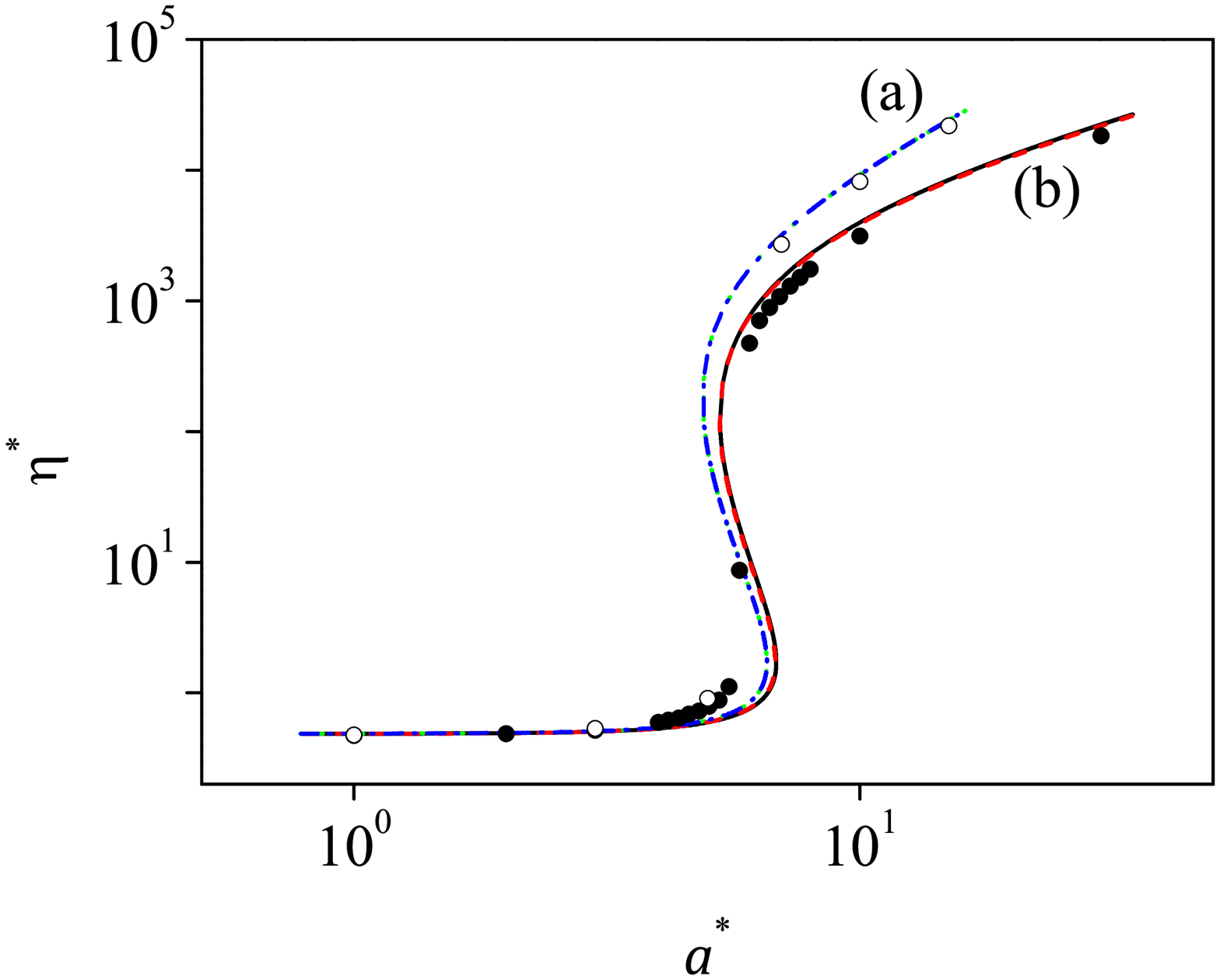}}
{\includegraphics[width=0.4\columnwidth]{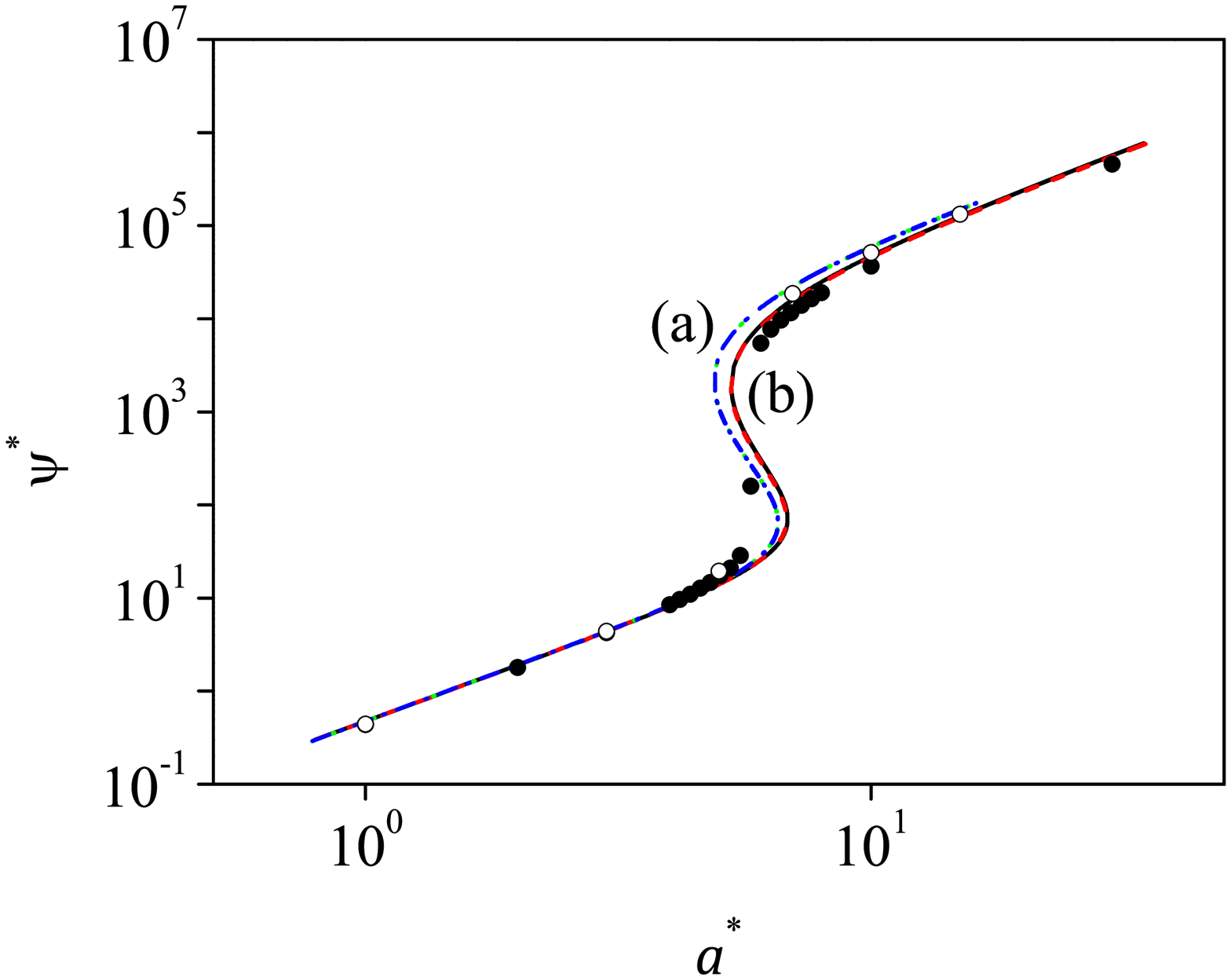}}
\caption{Plots of the steady granular temperature $\theta$, the non-Newtonian shear viscosity $\eta^*$, and the viscometric function $\Psi^*$ versus the (reduced) shear rate $a^*$ for $n^*=0.01$ and $T_\text{ex}^*=0.9$. Two different values of the coefficient of restitution $\al$ have been considered: $\al=1$ (a), and $\al=0.9$ (b). The solid and dotted lines correspond to the results obtained from the Boltzmann equation for IMM. The dashed and dash--dotted lines correspond to the results obtained from the BGK equation for IHS. Symbols refer to computer simulation results: empty circles for $\al=1$ and filled circles for $\al=0.9$.
\label{fig2}}
\end{figure}

In sections \ref{sec4} and \ref{sec5} we have solved the Boltzmann and BGK kinetic equations to obtain the shear-rate dependence of the second- and fourth-degree moments of a sheared granular suspension. In dimensionless form, those moments are given in terms of the coefficient of restitution $\al$, the reduced density $n^*\equiv n\sigma^d$, the (reduced) background temperature $T_\text{ex}^*\equiv T_\text{ex}/(m\sigma^2\gamma^2)$, and the (reduced) shear rate $a^*\equiv a/\gamma$. The theoretical results obtained for the steady (scaled) granular temperature $\theta$ and the rheological functions $\eta^*$ and $\Psi^*$ are compared here against recent event-driven simulations \cite{HTG17} performed for a three-dimensional system ($d=3$). In the simulations, $n^*=0.01$ and $T_\text{ex}^*=0.9$. Henceforth, we will consider these values for $n^*$ and $T_\text{ex}^*$ for the remaining plots displayed in this section.

The shear-rate dependence of $\theta$, $\eta^*$, and $\Psi^*$ is plotted in Fig.\ \ref{fig2} for two different values of the coefficient of restitution $\al$: $\al=1$ (elastic collisions) and $\al=0.9$ (inelastic collisions). The analytical expressions of the above quantities obtained from the Boltzmann equation for IMM are given by Eqs.\ \eqref{3.14}, \eqref{3.17}, and \eqref{3.18} while Eqs.\ \eqref{4.11}, \eqref{4.12}, and \eqref{4.14} correspond to the results derived from the BGK equation for IHS. Recall that the latter results coincide with those derived by solving the Boltzmann equation for IHS \cite{HTG17} via Grad's moment method \cite{G49}. First, it is quite apparent that the agreement of both theoretical results with simulations is excellent in the complete range of (scaled) shear rates analyzed. As in previous works on sheared granular flows \cite{G03}, the good agreement found here between IMM and simulations of IHS confirms again the reliability of IMM to reproduce the main trends observed for IHS. Moreover, as remarked in previous studies \cite{HT17,HTG17}, Fig.\ \ref{fig2} highlights the existence of a discontinuous shear thickening effect, namely, the non-Newtonian shear viscosity $\eta^*$ discontinuously increases/decreases (at a certain value of $a^*$) as the (scaled) shear rate gradually increases/decreases. The origin of this saddle-node bifurcation is a consequence of the connection between the behaviors of the non-Newtonian shear viscosity for small [Newtonian branch, Eq.\ \eqref{3.19}] and large [Bagnoldian branch, Eqs.\ \eqref{3.21} and \eqref{3.22}] shear rates. \vicente{At a more quantitative level, in the case of the viscosity $\eta^*$, we also observe that simulation data suggest a sharper transition than the one obtained from the analytical results. These discrepancies (which are qualitatively small) could be in part due to the limitations of the molecular chaos ansatz of the Boltzmann equation which are of course avoided in the molecular dynamics method.}

\vicente{It must be remarked that}  the results (both theory and simulations) reported in Ref.\ \cite{HTG17} have shown that there is a transition from discontinuous shear thickening in dilute suspensions to continuous shear thickening at relatively low density. This finding is consistent with previous works \cite{TK95,SMTK96} where only the transition between the quenched and the ignited states for the steady temperature $\theta$ was analyzed but it contrasts with typical experimental observations in dense suspensions. With respect to the impact of the coefficient of restitution $\al$ on rheology, we see that the effect of $\al$ on the viscometric function $\Psi^*$ is smaller than the one found for the temperature $\theta$ and the shear viscosity $\eta^*$.

\begin{figure}
{\includegraphics[width=0.4\columnwidth]{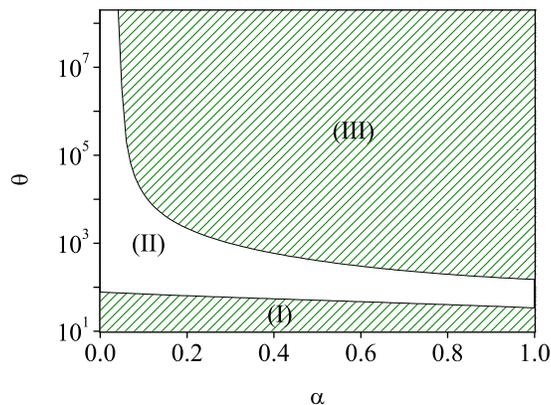}}
\caption{Phase diagram for the behavior of the (symmetric) fourth-degree moments for IMM. The hatched regions below the curve $\theta_c^{(1)}(\al)$ (region I) and above the curve $\theta_c^{(2)}(\al)$ (region III) correspond to states with well-defined values of the scaled fourth-degree moments. The region II [$\theta_c^{(1)}(\al)<\theta<\theta_c^{(2)}(\al)$] defines the states where  the fourth-degree moments have unphysical values. Here, $n^*=0.01$ and $T_\text{ex}^*=0.9$.
\label{fig3}}
\end{figure}
\begin{figure}
{\includegraphics[width=0.4\columnwidth]{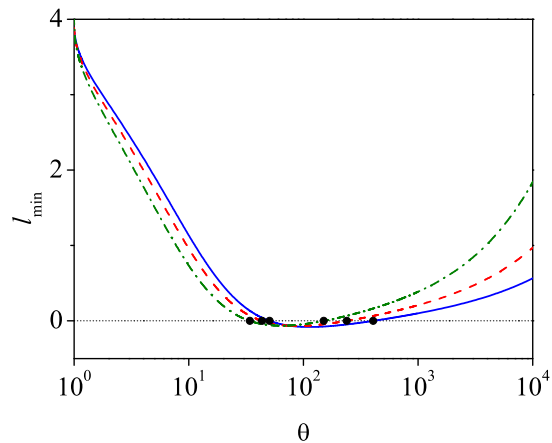}}
\caption{Plot of the smallest eigenvalue, $\ell_\text{min}$, associated with the time evolution of the (symmetric) fourth-degree moments for IMM as a function of the (scaled) temperature $\theta$ for $\al=0.5$ (solid line), $\al=0.7$ (dashed line), and $\al=1$ (dash-dotted line). The circles indicate the location of the corresponding values of the critical temperatures $\theta_c^{(1)}(\al)$ and $\theta_c^{(2)}(\al)$. Here, $n^*=0.01$ and $T_\text{ex}^*=0.9$.
\label{fig4}}
\end{figure}

We consider now the (symmetric) fourth-degree moments. They are given by Eq.\ \eqref{3.39} for the Boltzmann equation for IMM and Eq.\ \eqref{4.8} for the BGK kinetic model for IHS. As Fig.\ \ref{fig2} shows, the function $\theta(a^*)$ becomes a multi-valued function in a certain interval (in the vicinity of the saddle point) of values of the shear rate, namely, in this region there are two or three different values of $\theta$ leading to the same value of $a^*$. Thus, in order to detect the possible singularities of the fourth-degree moments, it is more convenient to use $\theta$ as input parameter instead of the (scaled) shear rate $a^*$. Once $\theta$ is known, $a^*(\theta)$ can be easily determined from Eqs.\ \eqref{3.14} and \eqref{4.14} for IMM and the BGK model, respectively. An inspection of the (simple) BGK-forms of these moments shows that they are well defined functions of both $\al$ and $\theta$ for any value of the coefficient of restitution $\al$. However, as occurs in dry granular gases \cite{SG07}, for any given value of $\al$, the matrix $\boldsymbol{\mathcal{L}}$ becomes singular ($\det \boldsymbol{\mathcal{L}}=0$) for two certain ``critical'' values $\theta_c^{(1)}(\al)$ and $\theta_c^{(2)}(\al)$, where $\theta_c^{(2)}(\al)>\theta_c^{(1)}(\al)$. This means that the (symmetric) fourth-degree moments tend to infinity when $\theta \to \theta_c^{(i)}$ $(i=1,2)$. Moreover, for $\theta_c^{(1)}(\al)<\theta<\theta_c^{(2)}(\al)$, the solutions to Eq.\ \eqref{3.39} are unphysical (e.g., $M_{4|0}^*<0$) and hence, the stationary USF is limited to the regions $0<\theta<\theta_c^{(1)}(\al)$ and $\theta>\theta_c^{(2)}(\al)$. The phase diagram associated with the singular behavior of the fourth-degree moments is plotted in Fig.\ \ref{fig3} for $n^*=0.01$ and $T_\text{ex}^*=0.9$. The curves $\theta_c^{(1)}(\al)$ (bottom curve) and $\theta_c^{(2)}(\al)$ (top curve) split the parameter space in three regions: the regions I and III correspond to states $(\theta,\al)$ with finite values of the fourth-degree moments while the region II defines the states where those moments have no physical values. Figure \ref{fig3} highlights the fact that the boundaries of the region II are nontrivial since at a given value of $\al$ there is a reentrance feature: we first find a transition from the region I (where the moments are well defined) to region II (unphysical values) by increasing the temperature $\theta$, followed by a subsequent transition to a well defined region (the region III). Moreover, while $\theta_c^{(2)}(\al)>\theta_c^{(1)}(\al)$, $a_c^{(1)*}(\al)>a_c^{(2)*}(\al)$ where $a_c^{(i)*}$ denotes the critical shear rate associated with $\theta_c^{(i)}$. As said before, $a_c^{(i)*}$ is determined from Eq.\ \eqref{3.14} by the replacement $\theta\to \theta_c^{(i)}$. As an example, at $\al=0.7$, $\theta_c^{(1)}=43.573$ and $\theta_c^{(2)}=238.639$ while $a_c^{(1)*}=7.437$ and $a_c^{(2)*}=6.441$. Similar behaviors are found for other values of $\al$.

It is important to recall that the divergence of the fourth-degree moments of the USF is also present for both elastic \cite{SGBD93,SG95,GS03} and inelastic \cite{SG07} Maxwell models. In both cases, an analysis of the time evolution of the fourth-degree moments shows that the eigenvalue $\ell_\text{min}$ of the matrix $\boldsymbol{\mathcal{L}}$ with the smallest real part governing the long time behavior of those moments becomes negative for shear rates larger than a critical value. Consequently, those moments exponentially grow in time (and so, they diverge in time) for $a^*>a_c^*$. To check if actually the origin of the singular behavior of the fourth-degree moments found here for granular suspensions is linked to the change of sign of the eigenvalue $\ell_\text{min}$, Fig.\ \ref{fig4} shows the dependence of $\ell_\text{min}$ on the (scaled) temperature $\theta$ for three different values of $\al$. At a given value of $\al$, we observe that $\ell_\text{min}$ exhibits a non-monotonic dependence on $\theta$ since it first decreases with increasing $\theta$, then it becomes negative in the region $\theta_c^{(1)}(\al)<\theta<\theta_c^{(2)}(\al)$, and eventually becomes positive for $\theta>\theta_c^{(2)}$ where it increases with increasing $\theta$. The corresponding critical values $\theta_c^{(1)}$ and $\theta_c^{(2)}$ are the same as those obtained from the condition $\det \boldsymbol{\mathcal{L}}=0$, confirming the above expectation.

\begin{figure}
{\includegraphics[width=0.4\columnwidth]{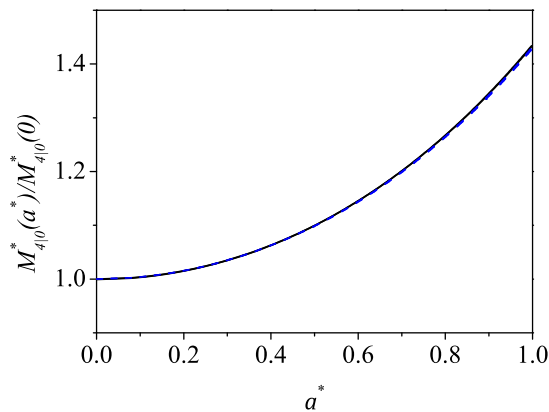}}
\caption{Plot of the scaled moment $M_{4|0}^*(a^*)/M_{4|0}^*(0)$ as a function of $a^*$ for $\al=0.7$ (solid and dotted lines) and 1 (dashed and dash-dotted lines). The solid and dashed lines correspond to the results obtained from the Boltzmann equation for IMM while the (indistinguishable) dotted and dash-dotted lines refer to the results obtained from the BGK equation for IHS.
Here, $n^*=0.01$ and $T_\text{ex}^*=0.9$.
\label{fig5}}
\end{figure}
\begin{figure}
{\includegraphics[width=0.4\columnwidth]{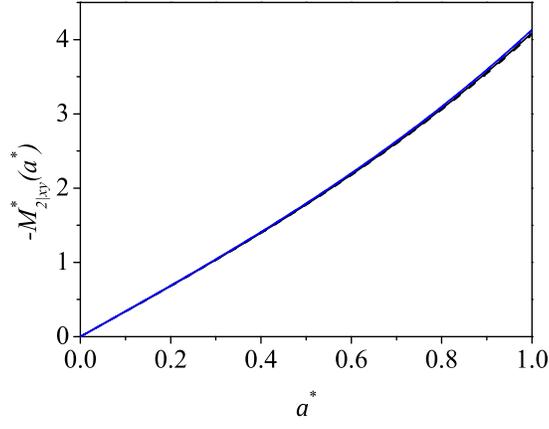}}
\caption{Plot of the scaled moment $-M_{2|xy}^*(a^*)$ as a function of $a^*$ for $\al=0.7$ (solid and dotted lines) and 1 (dashed and dash-dotted lines). The solid and dashed lines correspond to the results obtained from the Boltzmann equation for IMM while the (indistinguishable) dotted and dash-dotted lines refer to the results obtained from the BGK equation for IHS.
Here, $n^*=0.01$ and $T_\text{ex}^*=0.9$.
\label{fig6}}
\end{figure}

On the other hand, for states with $\theta<\theta_c^{(1)}(\al)$ and $\theta>\theta_c^{(2)}(\al)$ the (symmetric) fourth-degree moments have well-defined values and hence, one can study their shear-rate dependence. Here, for the sake of illustration, we consider the region $0<a^*<1$ where all the moments are well defined functions of the shear rate and in addition, nonlinear effects are still significant. Figure \ref{fig5} shows the ratio $M_{4|0}^*(a^*)/M_{4|0}^*(0)$ versus $a^*$ for $\al=1$ and 0.7. The results obtained for IMM from the Boltzmann equation are compared against the results derived for IHS from the BGK equation. This figure highlights that both theories agree perfectly well each other, even for quite relatively high values of the shear rate. Regarding the influence of collisional dissipation, we observe that the effect of $\al$ on the the moment $M_{4|0}^*$ is very tiny since all the results collapse in a common curve. It is appealing to remark the good performance of the BGK theoretical predictions for granular suspensions since previous comparisons \cite{GS95a} made for ordinary gases at the level of the fourth-degree moments have shown significant discrepancies between the Boltzmann (obtained for Maxwell molecules) and BGK results for large shear rates (say, $a^*\geq 0.2$). This disagreement is especially important for moments in which the component $V_x$ is the most relevant one. As a complement of Fig.\ \ref{fig5}, Fig.\ \ref{fig6} shows the shear-rate dependence of the magnitude of the (reduced) moment $M_{2|xy}^*$. This moment vanishes in the absence of shear rate ($a^*=0$). Similar conclusions to those made for the moment $M_{4|0}^*$ can be done for the moment $M_{2|xy}^*$.

\begin{figure}
{\includegraphics[width=0.4\columnwidth]{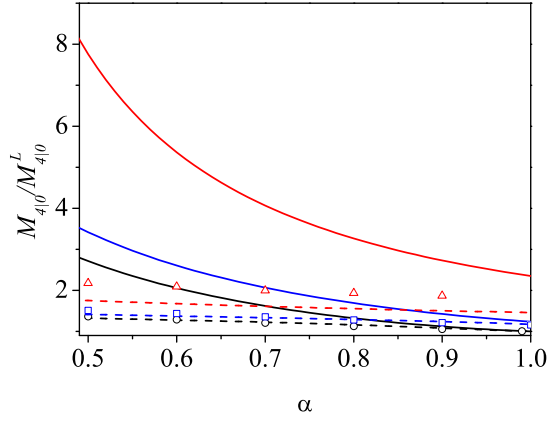}}
\caption{Plot of the reduced moment $M_{4|0}/M_{4|0}^{\text{L}}$ as a function of the coefficient of restitution $\al$ for three different values of the (reduced) friction coefficient $\widetilde{\gamma}=\gamma/\nu_0$: $\widetilde{\gamma}=0$ (black lines and circles),
$\widetilde{\gamma}=0.1$ (blue lines and squares), and $\widetilde{\gamma}=0.5$ (red lines and triangles).
The solid lines correspond to the results obtained from the Boltzmann equation for IMM while the dashed lines refer to the results derived from the BGK equation for IHS. Symbols refer to computer simulation results obtained in Ref.\ \cite{ChVG15}.
\label{fig7}}
\end{figure}
\begin{figure}
{\includegraphics[width=0.4\columnwidth]{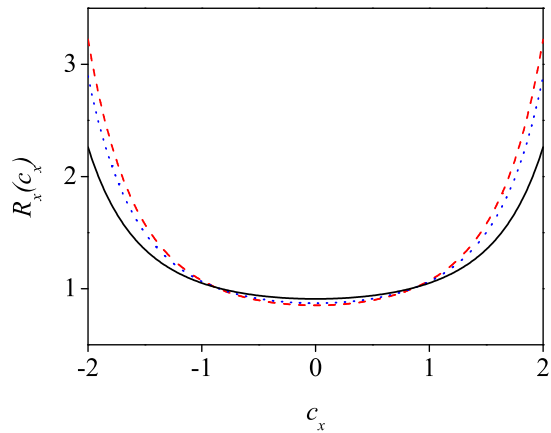}}
\caption{Plot of the ratio $R_x(c_x)=\varphi_x(c_x)/(\pi^{-1/2}e^{-c_x^2})$ versus the (scaled) velocity $c_x$ for $\widetilde{\gamma}=0.1$ and three different values of the coefficient of restitution $\al$: $\al=1$ (solid line), $\al=0.7$ (dotted line), and $\al=0.5$ (dashed line).
\label{fig8}}
\end{figure}

We consider now the special limit case $T_\text{ex}=0$ where computer simulations for the moment $M_{4|0}$ are available in the literature \cite{ChVG15}. In this limit case, the (reduced) shear rate $a^*$ is a function of the coefficient of restitution $\al$. Moreover, the (reduced) parameter $\widetilde{\gamma}\equiv \gamma/\nu_0$ is employed as input parameter in the DSMC results reported in Ref.\ \cite{ChVG15} instead of the background temperature $T_{\text{ex}}^*$. Figure \ref{fig7} shows the ratio $M_{4|0}/M_{4|0}^{\text{L}}$ versus $\al$ for three different values of $\widetilde{\gamma}$. Here,
\beq
\label{5.1}
M_{4|0}^{\text{L}}=\int\; \dd \mathbf{V} V^4 f_\text{\text{L}}(\mathbf{V}),
\eeq
where $f_\text{\text{L}}$ is defined in Eq.\ \eqref{4.2}. The solid and dashed lines refer to the results obtained from the Boltzmann equation for IMM and from the BGK equation for IHS, respectively. Symbols correspond to the computer simulation results obtained by numerically solving the Boltzmann equation for IHS by means of the DSMC method \cite{B94}. In the case of low values of the (reduced) friction coefficient $\widetilde{\gamma}$, we see that while the BGK results agree well with simulations in the full range of values of $\al$ represented here, more significant discrepancies between theory and simulations appear for IMM. On the other hand, the agreement between the BGK results and simulations is only qualitative for higher values of $\widetilde{\gamma}$ since the BGK predictions clearly underestimate the simulation results. Finally, Fig.\ \ref{fig8} plots the ratio $R_x(c_x)=\varphi_x(c_x)/(\pi^{-1/2}e^{-c_x^2})$ for $\widetilde{\gamma}=0.1$ and three different values of the coefficient of restitution $\al$. Here, the marginal distribution function $\varphi_x(c_x)$ is defined as
\beqa
\label{5.2}
\varphi_x(c_x)&=&\int_{-\infty}^{\infty}\; \dd c_y\;\int_{-\infty}^{\infty}\; \dd c_z\; \varphi(\mathbf{c})\nonumber\\
&=&\frac{1}{\sqrt{\pi}}\int_0^{\infty}\; \dd s \frac{e^{-(1-\widetilde{\lambda})s}}{\sqrt{1+\widetilde{a}^2s^2}}\text{exp}
\left(-e^{2\widetilde{\lambda}s} \frac{c_x^2}{1+\widetilde{a}^2s^2}\right),
\eeqa
where the scaled distribution $\varphi(\mathbf{c})$ is given by Eq.\ \eqref{4.26}. It is quite apparent that the distortion from equilibrium ($R_x\neq 1$) is more significant as the inelasticity increases. Although not shown here, comparison between theory and simulations (see Figs.\ 7 and 8 of Ref.\ \cite{ChVG15}) shows that while the BGK solution agrees very well with simulation data in the region of thermal velocities ($|c_x|\thicksim 1$), it exhibits quantitative discrepancies with simulations for larger velocities and strong collisional dissipation.

\section{Concluding remarks}
\label{sec6}

In spite of the simplicity of the USF, this state has been widely studied to shed light on the non-linear response of the system to strong shear rates. This response is accounted for by non-Newtonian transport properties such as the (scaled) temperature $\theta$, the (reduced) nonlinear shear viscosity $\eta^*$, and the (reduced) viscometric function $\Psi^*$. These properties are related to the second-degree velocity moments (pressure tensor). An interesting feature in \emph{sheared} granular suspensions (not shared by dry granular gases) is the so-called \emph{discontinuous} shear thickening effect, namely, the flow curve $\eta^*(a^*)$ has an $S$-shape, $a^*$ being the (reduced) shear rate. This means that, at a certain value of the shear rate, $\eta^*$ discontinuously increases/decreases if $a^*$ is gradually increased/decreased. This phenomena has been usually observed in dense systems and (apart from other factors) it has been recognized that the mutual friction between grains (\vicente{rough inelastic hard spheres}) plays an important role \cite{OH11,CPNC11,H13,SMMD13}. On the other hand, a more recent study \cite{HTG17} based on the Enskog kinetic equation has shown that the discontinuous shear thickening can be also found for smooth IHS in the dilute regime. The theoretical predictions for the rheological properties (which were obtained from Grad's moment method) were shown to compare very well with computer simulations, even for moderate densities. On the other hand, although the momentum transport is the most relevant phenomenon in a sheared suspension, higher degree moments are also important since they provide an indirect information of the velocity distribution function.

Given the intricacies embodied in the hard sphere kernel of the Boltzmann collision operator, to study the above issue one has to consider simplified collision models where velocity moments can be obtained without having to use approximate methods. In the context of the Boltzmann equation, the inelastic Maxwell model (IMM) allows us to determine higher-degree moments in the USF problem. In particular, the fourth-degree moments have been \emph{exactly} determined  for dry IMM \cite{GS07,SG07}. An appealing problem is to extend the previous efforts to the case of granular suspensions, namely, when the effect of the interstitial gas phase on solid particles is accounted for. This has likely been one of the main goals of the present contribution. In addition, to complement the results derived from the Boltzmann equation for IMM, a BGK-type kinetic model for granular suspensions \cite{BDS99} has been also solved to get all the velocity moments of the velocity distribution function.

\begin{figure}
{\includegraphics[width=0.4\columnwidth]{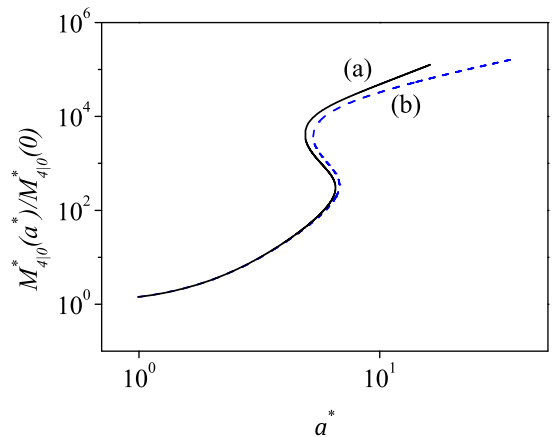}}
\caption{Shear-rate dependence of the scaled moment $M_{4|0}^*(a^*)/M_{4|0}^*(0)$ for $\al=1$ (solid line, (a)) and 0.9 (dashed line, (b)). The results are obtained from the BGK equation for IHS. Here, $n^*=0.01$ and $T_\text{ex}^*=0.9$.
\label{fig9}}
\end{figure}

As mentioned in the Introduction section, the motivation of our work is twofold. First, the comparison between the theoretical predictions for $\theta$, $\eta^*$, and $\Psi^*$ with computer simulations allow us to assess the accuracy of both approaches (IMM and BGK results) in conditions of practical interest. Thus, the results displayed in Fig.\ \ref{fig2} highlight the excellent performance of both theories in reproducing the shear-rate dependence of the rheological properties. \vicente{In particular, the exact results derived from the Boltzmann equation for IMM shows the existence of the so-called discontinuous shear thickening behavior where several mechanisms \cite{BJ09} have been proposed in the literature to explain the origin of this behavior. What is interesting here is the existence of this shear thickening in a \emph{structurally simple} system. In this case, these non-Newtonian properties are associated with both the behavior of the granular suspension in far from equilibrium situations as well as the impact of the interstitial fluid on the dynamics properties of the granular gas.} As a second aspect, the determination of the fourth-degree moments provides information on the combined effect of both the (reduced) shear rate and inelasticity on the high velocity population. In particular, an important result is that, for a given value of the coefficient of restitution $\al$, the (symmetric) fourth-degree moments of IMM have unphysical values in a certain region of the parameter space of the system. This singular behavior contrasts with the BGK results where all velocity moments are regular functions of both $a^*$ and $\al$. Since $\theta (a^*)$ is a multi-valued function (i.e., two or three values of $\theta$ correspond to the same value of $a^*$ for a certain range of values of $a^*$), it is more convenient to carry out the study on the divergence of the fourth-degree moments of IMM taking $\theta$ as independent parameter (input) instead of $a^*$. In this case, our results show that those moments are not well-defined in the region $\theta_c^{(1)}(\al)<\theta<\theta_c^{(2)}(\al)$ where the critical values $\theta_c^{(i)}(\al)$ are obtained from the condition $\det \boldsymbol{\mathcal L}=0$, where the matrix $\boldsymbol{\mathcal L}$ is defined by Eqs.\ \eqref{3.29} and \eqref{3.30}. Although this singularity of the fourth-degree moments for IMM is also present in elastic \cite{SGBD93,SG95} and inelastic \cite{SG07} systems, the phase diagram showing the regions where those moments are finite in granular suspensions is completely different to the one previously found for the above systems.

On the other hand, for states $\theta<\theta_c^{(1)}$ and $\theta>\theta_c^{(2)}$, the fourth-degree moments of IMM are well-defined functions. In particular, a comparison between the BGK and IMM results for those moments in the region $0\leq a^* \leq 1$ (where non-Newtonian effects are still important) surprisingly shows an excellent agreement between both theoretical results (see, for instance, Figs.\ \ref{fig5} and \ref{fig6}). This good performance of the BGK model contrasts with a previous comparison made for elastic Maxwell molecules \cite{GS95a} where the BGK predictions differ appreciably from the Boltzmann results for not too large shear rates (say, for instance, $a^*\gtrsim 0.2$). In addition, the shear-rate dependence of the fourth-degree moments is practically independent of inelasticity. It would be interesting to perform computer simulations to assess the accuracy of the above theoretical predictions for the fourth-degree moments.

\vicente{Although most of the previous works have focused on the study of discontinuous shear thickening effect of the non-Newtonian shear viscosity, a natural question is to see if actually the above behavior is also present in the fourth-degree moments. Since the BGK moments are well defined functions of both the coefficient of restitution and the shear rate, one may analyze the shear-rate dependence of those moments for high values of $a^*$. As an illustration, Fig.\ \ref{fig9} shows the scaled moment $M_{4|0}(a^*)/M_{4|0}(0)$ versus $a^*$ for $\al=1$ and 0.9. It is quite apparent that $M_{4|0}(a^*)/M_{4|0}(0)$ exhibits an $S$-shape since, at a given value of the shear rate, a small change in the shear rate produces a drastic increase of the fourth-degree moment $M_{4|0}^*$. This behavior has been also observed in the remaining (symmetric) fourth-degree moments. We expect that this theoretical prediction of the BGK model encourages the development of computer simulations to confirm this interesting result.}

As in many previous studies on granular gases, in this paper we have assumed that the coefficient of restitution $\al$ is a positive constant. It is well known that experimental observations \cite{BHL84} have shown that $\al$ depends on the impact velocity. The simplest model accounting for this velocity dependence of $\al$ is the model of viscoelastic particles \cite{BP00,BP03,DBPB13}. A possible extension of the results presented here along this direction could be an interesting problem. However, given that the discontinuous shear thickening for elastic suspensions is qualitatively similar to that of inelastic suspensions, we guess that the effect of the velocity dependence of $\al$ on the above phenomenon would be irrelevant. Another possible project would be \vicente{to consider the model of inelastic rough spheres \cite{GS95,Z06} where apart from the coefficient of normal restitution, a \emph{constant} coefficient of tangential restitution is introduced. This is a more realistic model than the model of smooth inelastic hard sphere since the inelasticity of collisions not only affects to the translational degrees of freedom but also to the rotational ones. The extension of the present results to this model would allow us to assess the impact of roughness on the discontinuous shear thickening problem.} Finally, it would be also appealing to study the case of multicomponent granular suspensions where problems like segregation can be addressed. Work along these lines are underway.

\acknowledgments

We want to thank Satoshi Takada and Mois\'es Garc\'ia Chamorro for providing us the simulation data included in Figs.\ \ref{fig2} and \ref{fig7}, respectively. The present work has been supported by the Spanish Government through Grant No. FIS2016-76359-P, partially financed by ôFondo Europeo de Desarrollo Regionalö funds. The research of Rub\'en G\'omez Gonz\'alez has been supported by the predoctoral fellowship BES-2017-079725 from the Spanish Government.


\appendix

\section{Fourth-degree collisional moments of IMM}
\label{appA}

In this Appendix, the expressions of the relevant fourth-degree collisional moments in a three-dimensional system are displayed. The explicit forms of these moments were obtained in Ref.\ \cite{GS07}. As mentioned in section \ref{sec3}, there are eight independent \emph{symmetric} (or nonvanishing) moments in the geometry of the steady USF state. They are given by the set
\beq
\label{a3}
\left\{M_{4|0}, M_{2|xx}, M_{2|yy}, M_{2|xy}, M_{0|yyyy}, M_{0|zzzz}, M_{0|xxxy}, M_{0|yyyx}\right\}.
\eeq
where the moments $M_{4|0}$, $M_{2|ij}$, and $M_{0|ijk\ell}$ are defined by Eq. \eqref{a4}. Their corresponding collisional moments are given by
\beq
\label{a5}
\Big(J_{4|0}, J_{2|ij}, J_{0|ijk\ell}\Big)=\int \dd\mathbf{V}\; \Big(Y_{4|0}, Y_{2|ij}, Y_{0|ijk\ell}\Big) J_\text{IMM}[\mathbf{V}|f,f].
\eeq
The explicit expressions for the collisional moments are \cite{GS07}
\beq
\label{a6}
J_{4|0}=-\nu_{4|0}M_{4|0}+9\frac{p^2}{n m^2}\lambda_1-\frac{\lambda_2}{n m^2}\Pi_{k\ell}\Pi_{k\ell},
\eeq
\beq
\label{a7}
J_{2|xx}=-\nu_{2|2}M_{2|xx}+3\lambda_3\frac{p}{n m^2}\Pi_{xx}-\frac{\lambda_4}{n m^2}\left(\Pi_{xk}\Pi_{kx}-\frac{1}{3}\Pi_{k\ell}\Pi_{\ell k}\right),
\eeq
\beq
\label{a8}
J_{2|xy}=-\nu_{2|2}M_{2|xy}+3\lambda_3\frac{p^2}{n m^2}\Pi_{xy}-\frac{\lambda_4}{n m^2}\Pi_{xk}\Pi_{ky},
\eeq
\beq
\label{a9}
J_{0|yyyy}=-\nu_{0|4}M_{0|yyyy}+3\frac{\lambda_5}{n m^2}\left(\Pi_{yy}^2-\frac{4}{7}\Pi_{yk}\Pi_{ky}+\frac{2}{35}\Pi_{k\ell}\Pi_{\ell k}\right),
\eeq
\beq
\label{a10}
J_{0|xxxy}=-\nu_{0|4}M_{0|xxxy}+3\frac{\lambda_5}{n m^2}\left(\Pi_{xx}\Pi_{yy}-\frac{2}{7}\Pi_{xk}\Pi_{ky}\right).
\eeq
The collisional moments $J_{2|yy}$, $J_{0|zzzz}$, and $J_{0|yyyx}$ can be easily obtained from Eqs.\ \eqref{a7}, \eqref{a9}, and \eqref{a10}, respectively. In Eqs.\ \eqref{a6}--\eqref{a10}, the usual Einstein summation convention over repeated indices is assumed. Moreover, we have introduced the effective collision frequencies
\beq
\label{a11}
\nu_{4|0}=2\zeta+\frac{(1+\al)^2(5+6\al-3\al^2)}{120}\nu_\text{M}, \quad \nu_{2|2}=2\zeta+\frac{(1+\al)^2(34+21\al-6\al^2)}{420}\nu_\text{M},
\eeq
\beq
\label{a12}
\nu_{0|4}=2\zeta+\frac{(1+\al)^2(50+7\al-\al^2)}{315}\nu_\text{M},
\eeq
where $\zeta=(1-\al^2)\nu_\text{M}/6$. Finally, the cross coefficients $\lambda_i$ in Eqs.\ \eqref{a6}--\eqref{a10} are given by
\beq
\label{a13}
\lambda_1=\frac{(1+\al)^2(11-6\al+3\al^2)}{72}\nu_\text{M}, \quad \lambda_2=\frac{(1+\al)^2(1+6\al-3\al^2)}{60}\nu_\text{M},
\quad \lambda_3=\frac{(1+\al)^2(22-21\al+6\al^2)}{180}\nu_\text{M},
\eeq
\beq
\label{a14}
\lambda_4=\frac{(1+\al)^2(21\al-3\al^2-1)}{210}\nu_\text{M}, \quad \lambda_5=\frac{(1+\al)^2(39-21\al+3\al^2-1)}{945}\nu_\text{M}.
\eeq

\section{Results from the BGK-type kinetic model}
\label{appB}

The results derived from the BGK kinetic model are displayed in this Appendix. Let us consider first Eq.\ \eqref{4.4}:
\beq
\label{b1}
a k_1 M_{k_1-1,k_2+1,k_3}+\left(\chi \nu_0+k \lambda\right)M_{k_1,k_2,k_3}=N_{k_1,k_2,k_3},
\eeq
where $N_{k_1,k_2,k_3}$ is defined by Eq.\ \eqref{4.5}. Given that $R_{k_1,k_2,k_3}$ is a linear combination of velocity moments of degree $k-2$, the quantity $N_{k_1,k_2,k_3}$ is assumed to be known in the equation defining the moments $M_{k_1,k_2,k_3}$ of degree $k$. To solve the hierarchy of moment equations \eqref{b1}, we introduce the operators $L_1$ and $L_2$ acting on functions $\psi(k_1,k_2,k_3)$ as
\beq
\label{b2}
L_1 \psi(k_1,k_2,k_3)=\psi(k_1-1,k_2,k_3), \quad  L_2 \psi(k_1,k_2,k_3)=\psi(k_1,k_2+1,k_3).
\eeq
Thus, Eq.\ \eqref{b1} can be written as
\beq
\label{b3}
\left(a k_1 L_1L_2+\chi \nu_0+k \lambda\right)M_{k_1,k_2,k_3}=N_{k_1,k_2,k_3}.
\eeq
Its formal solution is
\beq
\label{b4}
M_{k_1,k_2,k_3}=\left(a k_1 L_1L_2+\chi \nu_0+k \lambda\right)^{-1}N_{k_1,k_2,k_3}.
\eeq
Since
\beq
\label{b5}
L_1 L_2 \left[\chi \nu+(k_1+k_2+k_3) \lambda\right]=\chi \nu_0+(k_1+k_2+k_3) \lambda,
\eeq
then, the solution \eqref{b4} can be written more explicitly as
\beqa
\label{b6}
M_{k_1,k_2,k_3}&=&\frac{1}{\chi \nu_0+k \lambda}\left(1+\frac{a k_1}{\chi \nu_0+k \lambda}L_1L_2\right)^{-1}N_{k_1,k_2,k_3}
\nonumber\\
&=&\sum_{q=0}^\infty \frac{(-a)^q}{\left(\chi \nu_0+k \lambda\right)^{1+q}}\left(k_1 L_1 L_2\right)^q N_{k_1,k_2,k_3}.
\eeqa
On the other hand, it is straightforward to prove that
\beq
\label{b7}
\left(k_1 L_1 L_2\right)^q N_{k_1,k_2,k_3}=\frac{k_1!}{(k_1-q)!}N_{k_1-q,k_2+q,k_3},
\eeq
if $q\leq k_1$, being zero otherwise. Thus, Eq.\ \eqref{b6} can be finally written in the form
\beq
\label{b8}
M_{k_1,k_2,k_3}=\sum_{q=0}^{k_1} \frac{k_1!}{(k_1-q)!} \frac{(-a)^q}{\left(\chi \nu_0+k \lambda\right)^{1+q}}N_{k_1-q,k_2+q,k_3}.
\eeq




\begin{thebibliography}{51}
\expandafter\ifx\csname natexlab\endcsname\relax\def\natexlab#1{#1}\fi
\expandafter\ifx\csname bibnamefont\endcsname\relax
  \def\bibnamefont#1{#1}\fi
\expandafter\ifx\csname bibfnamefont\endcsname\relax
  \def\bibfnamefont#1{#1}\fi
\expandafter\ifx\csname citenamefont\endcsname\relax
  \def\citenamefont#1{#1}\fi
\expandafter\ifx\csname url\endcsname\relax
  \def\url#1{\texttt{#1}}\fi
\expandafter\ifx\csname urlprefix\endcsname\relax\def\urlprefix{URL }\fi
\providecommand{\bibinfo}[2]{#2}
\providecommand{\eprint}[2][]{\url{#2}}


\bibitem[{\citenamefont{Brown and Jeager}(2014)}]{BJ14}
\bibinfo{author}{\bibnamefont{Brown} E} \bibnamefont{and}
  \bibinfo{author}{\bibfnamefont{H.~M.} \bibnamefont{Jeager} H M}, 2014
  \bibinfo{journal}{Rep. Prog. Phys.} \textbf{\bibinfo{volume}{77}},
  \bibinfo{pages}{046602}


\bibitem[{\citenamefont{Barnes}(1989)}]{B89}
\bibinfo{author}{\bibnamefont{Barnes} \bibfnamefont{H~A} }, 1989
  \bibinfo{journal}{J. Rheol.} \textbf{\bibinfo{volume}{33}},
  \bibinfo{pages}{329}

\bibitem[{\citenamefont{Lootens et~al.}(2005)\citenamefont{Lootens, van Damme,
  H\'emar, and H\'ebraud}}]{LDHH05}
\bibinfo{author}{\bibnamefont{Lootens} D},
  \bibinfo{author}{\bibnamefont{van Damme} H},
  \bibinfo{author}{\bibnamefont{H\'emar} Y} \bibnamefont{and}
  \bibinfo{author}{\bibnamefont{H\'ebraud} P}, 2005
  \bibinfo{journal}{Phys. Rev. Lett.} \textbf{\bibinfo{volume}{95}},
  \bibinfo{pages}{268302}

\bibitem[{\citenamefont{Brown and Jaeger}(2009)}]{BJ09}
\bibinfo{author}{\bibnamefont{Brown} E} \bibnamefont{and}
  \bibinfo{author}{\bibnamefont{Jaeger} H M}, 2009
  \bibinfo{journal}{Phys. Rev. Lett.} \textbf{\bibinfo{volume}{103}},
  \bibinfo{pages}{086001}

\bibitem[{\citenamefont{Mewis and Wagner}(2011)}]{MW11}
\bibinfo{author}{\bibnamefont{Mewis} J} \bibnamefont{and}
  \bibinfo{author}{ \bibnamefont{Wagner} \bibfnamefont{N~J}}, 2011
  \emph{\bibinfo{title}{Colloidal Suspension Rheology}}
  (\bibinfo{publisher}{Cambridge University Press: New York})

\bibitem[{\citenamefont{Ciamarra et~al.}(2011)\citenamefont{Ciamarra, Pastore,
  Nicodemi, and Coniglio}}]{CPNC11}
\bibinfo{author}{\bibnamefont{Ciamarra} \bibfnamefont{M~P}},
  \bibinfo{author}{\bibnamefont{Pastore} \bibfnamefont{R}},
  \bibinfo{author}{\bibnamefont{Nicodemi} M} \bibnamefont{and}
  \bibinfo{author}{\bibnamefont{Coniglio} A}, 2011
  \bibinfo{journal}{Phys. Rev. E} \textbf{\bibinfo{volume}{84}},
  \bibinfo{pages}{041308}

\bibitem[{\citenamefont{Otsuki and Hayakawa}(2011)}]{OH11}
\bibinfo{author}{\bibnamefont{Otsuki} M} \bibnamefont{and}
  \bibinfo{author}{\bibnamefont{Hayakawa} H}, 2011
  \bibinfo{journal}{Phys. Rev. E} \textbf{\bibinfo{volume}{83}},
  \bibinfo{pages}{051301}

\bibitem[{\citenamefont{Heussinger}(2013)}]{H13}
\bibinfo{author}{\bibnamefont{Heussinger} C}, 2013
  \bibinfo{journal}{Phys. Rev. E} \textbf{\bibinfo{volume}{88}},
  \bibinfo{pages}{{050}{201}(R)}

\bibitem[{\citenamefont{Seto et~al.}(2013)\citenamefont{Seto, Mari, Morris, and
  Denn}}]{SMMD13}
\bibinfo{author}{\bibnamefont{Seto} R},
  \bibinfo{author}{\bibnamefont{Mari} R},
  \bibinfo{author}{\bibnamefont{Morris} J F}
  \bibnamefont{and} \bibinfo{author}{\bibnamefont{Denn} M M}, 2013
  \bibinfo{journal}{Phys. Rev. Lett.} \textbf{\bibinfo{volume}{111}},
  \bibinfo{pages}{{218}{301}}



\bibitem[{\citenamefont{Kranz et~al.}(2018)\citenamefont{Kranz, Frahsa,
  Zippelius, Fuchs, and Sperl}}]{KFZFS18}
\bibinfo{author}{\bibnamefont{Kranz} W T},
  \bibinfo{author}{\bibnamefont{Frahsa} F},
  \bibinfo{author}{\bibnamefont{Zippelius} A},
  \bibinfo{author}{\bibnamefont{Fuchs} M} \bibnamefont{and}
  \bibinfo{author}{\bibnamefont{Sperl} M}, 2018
  \bibinfo{journal}{Phys. Rev. Lett.} \textbf{\bibinfo{volume}{121}},
  \bibinfo{pages}{148002}


\bibitem{BB85}Brady J F and Bossis G, 1985 J. Fluid Mech. \textbf{155}, 105

\bibitem{WB09}Wagner N J and Brady J F, 2009 Phys. Today \textbf{62}, 27

\bibitem{H74}Hoffman R L, 1974 J. Colloid Interface Sci. \textbf{46}, 491

\bibitem{H82}Hoffman R L, 1982 Adv. Colloid Interface Sci. \textbf{17}, 161

\bibitem{CHH05}Cates M E, Haw M D and Holmes C B, 2005 J. Phys: Condens. Matter \textbf{17} S2517

\bibitem{BJ12}Brown E and Jaeger H M, 2012 J. Rheol. \textbf{39}, 875


\bibitem[{\citenamefont{Tsao and Koch}(1995)}]{TK95}
\bibinfo{author}{\bibnamefont{Tsao} H-K} \bibnamefont{and}
  \bibinfo{author}{\bibnamefont{Koch} D}, 1995
  \bibinfo{journal}{J. Fluid Mech.} \textbf{\bibinfo{volume}{296}},
  \bibinfo{pages}{211}

\bibitem[{\citenamefont{Sangani et~al.}(1996)\citenamefont{Sangani, Mo, Tsao,
  and Koch}}]{SMTK96}
\bibinfo{author}{\bibnamefont{Sangani} A S},
  \bibinfo{author}{\bibnamefont{Mo} G},
  \bibinfo{author}{\bibnamefont{Tsao} H-K} \bibnamefont{and}
  \bibinfo{author}{\bibnamefont{Koch} D}, 1996
  \bibinfo{journal}{J. Fluid Mech.} \textbf{\bibinfo{volume}{313}},
  \bibinfo{pages}{309}

\bibitem{SA17}Saha S and Alam M, 2017 J. Fluid Mech. \textbf{833}, 206

\bibitem[{\citenamefont{Hayakawa et~al.}(2017)\citenamefont{Hayakawa, Takada,
  and Garz\'o}}]{HTG17}
\bibinfo{author}{\bibnamefont{Hayakawa} H},
  \bibinfo{author}{\bibnamefont{Takada} S} \bibnamefont{and}
  \bibinfo{author}{\bibnamefont{Garz\'o} V}, 2017
  \bibinfo{journal}{Phys. Rev. E} \textbf{\bibinfo{volume}{96}},
  \bibinfo{pages}{042903}

\bibitem[{\citenamefont{Cercignani}(1988)}]{C88}
\bibinfo{author}{\bibnamefont{Cercignani} C}, 1988
  \emph{\bibinfo{title}{The Boltzmann Equation and Its Applications}}
  (\bibinfo{publisher}{Springer--Verlag: New York})

\bibitem[{\citenamefont{Chapman and Cowling}(1970)}]{CC70}
\bibinfo{author}{\bibnamefont{Chapman} S} \bibnamefont{and}
  \bibinfo{author}{\bibnamefont{Cowling} T G}, 1970
  \emph{\bibinfo{title}{The Mathematical Theory of Nonuniform Gases}}
  (\bibinfo{publisher}{Cambridge University Press: Cambridge})

\bibitem[{\citenamefont{Ferziger and Kaper}(1972)}]{FK72}
\bibinfo{author}{\bibnamefont{Ferziger} J H} \bibnamefont{and}
  \bibinfo{author}{\bibnamefont{Kaper} G H}, 1972
  \emph{\bibinfo{title}{Mathematical Theory of Transport Processes in Gases}}
  (\bibinfo{publisher}{North--Holland: Amsterdam})

\bibitem[{\citenamefont{Garz\'o and Santos}(2003)}]{GS03}
\bibinfo{author}{\bibnamefont{Garz\'o} V}  \bibnamefont{and}
  \bibinfo{author}{\bibnamefont{Santos} \bibfnamefont{A}},
  \emph{\bibinfo{title}{Kinetic Theory of Gases in Shear Flows. Nonlinear
  Transport}} (\bibinfo{publisher}{Kluwer Academic Publishers, Dordrecht},
  \bibinfo{year}{2003}).

\bibitem[{\citenamefont{Brey et~al.}(1999)\citenamefont{Brey, Dufty, and
  Santos}}]{BDS99}
\bibinfo{author}{\bibnamefont{Brey} J J},
  \bibinfo{author}{\bibnamefont{Dufty} J W} \bibnamefont{and}
  \bibinfo{author}{\bibnamefont{Santos} A}, 1999
  \bibinfo{journal}{J. Stat. Phys.} \textbf{\bibinfo{volume}{97}},
  \bibinfo{pages}{281}

\bibitem[{\citenamefont{Santos et~al.}(1993)\citenamefont{Santos, Garz\'o,
  Brey, and Dufty}}]{SGBD93}
\bibinfo{author}{\bibnamefont{Santos} A},
  \bibinfo{author}{\bibnamefont{Garz\'o} V},
  \bibinfo{author}{\bibnamefont{Brey} J J} \bibnamefont{and}
  \bibinfo{author}{\bibnamefont{Dufty} J W}, 1993
  \bibinfo{journal}{Phys. Rev. Lett.} \textbf{\bibinfo{volume}{71}},
  \bibinfo{pages}{3971}\\
  \bibinfo{author}{\bibnamefont{Santos} A},
  \bibinfo{author}{\bibnamefont{Garz\'o} V},
  \bibinfo{author}{\bibnamefont{Brey} J J} \bibnamefont{and}
  \bibinfo{author}{\bibnamefont{Dufty} J W}, 1994
  \bibinfo{journal}{Phys. Rev. Lett.} \textbf{\bibinfo{volume}{72}},
  \bibinfo{pages}{1392} (erratum)

\bibitem[{\citenamefont{Santos and Garz\'o}(1995)}]{SG95}
\bibinfo{author}{\bibnamefont{Santos} A} \bibnamefont{and}
  \bibinfo{author}{\bibnamefont{Garz\'o} V}, 1995
  \bibinfo{journal}{Physica A} \textbf{\bibinfo{volume}{213}}

\bibitem[{\citenamefont{Santos and Garz\'o}(2007)}]{SG07}
\bibinfo{author}{\bibnamefont{Santos} A} \bibnamefont{and}
  \bibinfo{author}{\bibnamefont{Garz\'o} V}, 2007
  \bibinfo{journal}{J. Stat. Mech.} \textbf{\bibinfo{volume}{P08021}}


\bibitem[{\citenamefont{Garz\'o}(2007)}]{G07bis}
\bibinfo{author}{\bibnamefont{Garz\'o} V}, 2007 \bibinfo{journal}{J.
  Phys. A: Math. Theor.} \textbf{\bibinfo{volume}{40}},
  \bibinfo{pages}{{107}{29}}

\bibitem[{\citenamefont{Brilliantov and P\"oschel}(2004)}]{BP04}
\bibinfo{author}{\bibnamefont{Brilliantov} N} \bibnamefont{and}
  \bibinfo{author}{\bibnamefont{P\"oschel} T}, 2004
  \emph{\bibinfo{title}{Kinetic Theory of Granular Gases}}
  (\bibinfo{publisher}{Oxford University Press: Oxford}).

\bibitem[{\citenamefont{Garz\'o et~al.}(2012)\citenamefont{Garz\'o, Tenneti,
  Subramaniam, and Hrenya}}]{GTSH12}
\bibinfo{author}{\bibnamefont{Garz\'o} V},
  \bibinfo{author}{\bibnamefont{Tenneti} S},
  \bibinfo{author}{\bibnamefont{Subramaniam} S}
  \bibnamefont{and} \bibinfo{author}{
  \bibnamefont{Hrenya} C M}, 2012 \bibinfo{journal}{J. Fluid Mech.}
  \textbf{\bibinfo{volume}{712}}, \bibinfo{pages}{129}

\bibitem[{\citenamefont{Garz\'o et~al.}(2016)\citenamefont{Garz\'o, Fullmer,
  Hrenya, and Yin}}]{GFHY16}
\bibinfo{author}{\bibnamefont{Garz\'o} V},
  \bibinfo{author}{\bibnamefont{Fullmer} W D},
  \bibinfo{author}{\bibnamefont{Hrenya} C M}
  \bibnamefont{and} \bibinfo{author}{\bibnamefont{Yin} X}, 2016
  \bibinfo{journal}{Phys. Rev. E} \textbf{\bibinfo{volume}{93}},
  \bibinfo{pages}{012905}

\bibitem[{\citenamefont{Williams and MacKintosh}(1996)}]{WM96}
\bibinfo{author}{\bibnamefont{Williams} D R M}
  \bibnamefont{and} \bibinfo{author}{
  \bibnamefont{MacKintosh} F C}, 1996 \bibinfo{journal}{Phys. Rev. E}
  \textbf{\bibinfo{volume}{54}}, \bibinfo{pages}{R9}

\bibitem[{\citenamefont{van Kampen}(1981)}]{K81}
\bibinfo{author}{\bibnamefont{van Kampen} N G}, 1981 \emph{\bibinfo{title}{Stochastic Processes in Physics and Chemistry}}
  (\bibinfo{publisher}{North--Holland: Amsterdam})

\bibitem[{\citenamefont{van Noije and Ernst}(1998)}]{NE98}
\bibinfo{author}{\bibnamefont{van Noije} T P C}
  \bibnamefont{and} \bibinfo{author}{\bibnamefont{Ernst} M H}, 1998
  \bibinfo{journal}{Granular Matter} \textbf{\bibinfo{volume}{1}},
  \bibinfo{pages}{57}

\bibitem[{\citenamefont{Koch and Hill}(2001)}]{KH01}
\bibinfo{author}{\bibnamefont{Koch} D L} \bibnamefont{and}
  \bibinfo{author}{\bibnamefont{Hill} R J}, 2001
  \bibinfo{journal}{Annu. Rev. Fluid Mech.} \textbf{\bibinfo{volume}{33}},
  \bibinfo{pages}{619}

\bibitem[{\citenamefont{Koch}(1990)}]{K90}
\bibinfo{author}{\bibnamefont{Koch} D L}, 1990
  \bibinfo{journal}{Phys. Fluids A} \textbf{\bibinfo{volume}{2}},
  \bibinfo{pages}{1711}

\bibitem[{\citenamefont{Wylie et~al.}(2003)\citenamefont{Wylie, Koch, and
  Ladd}}]{WKL03}
\bibinfo{author}{\bibnamefont{Wylie} J J},
  \bibinfo{author}{\bibnamefont{Koch} D L} \bibnamefont{and}
  \bibinfo{author}{\bibnamefont{Ladd} A J C}, 2003
  \bibinfo{journal}{J. Fluid Mech.} \textbf{\bibinfo{volume}{480}},
  \bibinfo{pages}{95}

\bibitem[{\citenamefont{Chamorro et~al.}(2015)\citenamefont{Chamorro,
  Vega~Reyes, and Garz\'o}}]{ChVG15}
\bibinfo{author}{\bibnamefont{Chamorro} M G},
  \bibinfo{author}{\bibnamefont{Vega~Reyes} F}
  \bibnamefont{and} \bibinfo{author}{\bibnamefont{Garz\'o} V}, 2015
  \bibinfo{journal}{Phys. Rev. E} \textbf{\bibinfo{volume}{92}},
  \bibinfo{pages}{{052}{205}}

\bibitem[{\citenamefont{Dufty et~al.}(1986)\citenamefont{Dufty, Santos, Brey,
  and Rodr\'{\i}guez}}]{DSBR86}
\bibinfo{author}{\bibnamefont{Dufty} J W},
  \bibinfo{author}{\bibnamefont{Santos} A},
  \bibinfo{author}{\bibnamefont{Brey} J J} \bibnamefont{and}
  \bibinfo{author}{\bibnamefont{Rodr\'{\i}guez} R F}, 1986
  \bibinfo{journal}{Phys. Rev. A} \textbf{\bibinfo{volume}{33}},
  \bibinfo{pages}{459}

\bibitem[{\citenamefont{Grad}(1949)}]{G49}
\bibinfo{author}{\bibnamefont{Grad} H}, 1949
  \bibinfo{journal}{Commun. Pure Appl. Math.} \textbf{\bibinfo{volume}{2}},
  \bibinfo{pages}{331}

\bibitem[{\citenamefont{Ben-Naim and Krapivsky}(2003)}]{BK03}
\bibinfo{author}{\bibnamefont{Ben-Naim} E} \bibnamefont{and}
  \bibinfo{author}{\bibnamefont{Krapivsky} P L}, 2003
  \emph{\bibinfo{booktitle}{Granular {G}as {D}ynamics} (Lectures Notes in Physics \emph{vol 624})}, ed
  \bibinfo{editor}{\bibfnamefont{T}~\bibnamefont{P\"oschel}} \bibnamefont{and}
  \bibinfo{editor}{\bibfnamefont{N}~\bibnamefont{Brilliantov}} (Berlin: Springer) pp 65--93


\bibitem[{\citenamefont{Garz\'o and Santos}(2007)}]{GS07}
\bibinfo{author}{\bibnamefont{Garz\'o} V} \bibnamefont{and}
  \bibinfo{author}{\bibnamefont{Santos} A}, 2007
  \bibinfo{journal}{J. Phys. A: Math. Theor.} \textbf{\bibinfo{volume}{40}},
  \bibinfo{pages}{{14}{927}}
  
  
  
\bibitem{SMDB98}Santos A, Montanero J M, Dufty J W and Brey J J, 1998 Phys. Rev. E \textbf{57}, 1644  

\bibitem[{\citenamefont{Santos}(2003)}]{S03}
\bibinfo{author}{\bibnamefont{Santos} A}, 2003
  \bibinfo{journal}{Physica A} \textbf{\bibinfo{volume}{321}},
  \bibinfo{pages}{442}

\bibitem[{\citenamefont{Truesdell and Muncaster}(1980)}]{TM80}
\bibinfo{author}{\bibnamefont{Truesdell} C} \bibnamefont{and}
  \bibinfo{author}{\bibnamefont{Muncaster} R G}, 1980
  \emph{\bibinfo{title}{Fundamentals of {M}axwell's {K}inetic {T}heory of a
  {S}imple {M}onatomic {G}as}} (\bibinfo{publisher}{Academic Press: New York})

\bibitem[{\citenamefont{Hayakawa and Takada}(2017)}]{HT17}
\bibinfo{author}{\bibnamefont{Hayakawa} H} \bibnamefont{and}
  \bibinfo{author}{\bibnamefont{Takada} S}, 2017
  \bibinfo{journal}{Powders\&Grains 2017, EPJ Web of Conferences}
  \textbf{\bibinfo{volume}{140}}, \bibinfo{pages}{09003}


\bibitem[{\citenamefont{Wylie and Koch}(2000)}]{WK00}
\bibinfo{author}{\bibnamefont{Wylie} J J} \bibnamefont{and}
  \bibinfo{author}{\bibnamefont{Koch} D L}, 2000
  \bibinfo{journal}{Phys. Fluids} \textbf{\bibinfo{volume}{12}},
  \bibinfo{pages}{964}

\bibitem[{\citenamefont{Wylie et~al.}(2009)\citenamefont{Wylie, Zhang, Li, and
  Hengyi}}]{WZLH09}
\bibinfo{author}{\bibnamefont{Wylie} J J},
  \bibinfo{author}{\bibnamefont{Zhang} Q},
  \bibinfo{author}{\bibnamefont{Li} Y} \bibnamefont{and}
  \bibinfo{author}{\bibnamefont{Hengyi} X}, 2009
  \bibinfo{journal}{Phys. Rev. E} \textbf{\bibinfo{volume}{79}},
  \bibinfo{pages}{031301}

\bibitem[{\citenamefont{Hilton and Tordesillas}(2013)}]{HT13}
\bibinfo{author}{\bibnamefont{Hilton} J E} \bibnamefont{and}
  \bibinfo{author}{\bibnamefont{Tordesillas} A}, 2013
  \bibinfo{journal}{Phys. Rev. E} \textbf{\bibinfo{volume}{88}},
  \bibinfo{pages}{062203}

\bibitem[{\citenamefont{Wang et~al.}(2014)\citenamefont{Wang, Grob, Zippelius,
  and Sperl}}]{WGZS14}
\bibinfo{author}{\bibnamefont{Wang} T},
  \bibinfo{author}{\bibnamefont{Grob} M},
  \bibinfo{author}{\bibnamefont{Zippelius} A}
  \bibnamefont{and} \bibinfo{author}{\bibnamefont{Sperl} M}, 2014
  \bibinfo{journal}{Phys. Rev. E} \textbf{\bibinfo{volume}{89}},
  \bibinfo{pages}{{042}{209}}
  
\bibitem{GSB90}Garz\'o V, Santos A and Brey J J, 1990 Physica A \textbf{163}, 651 

\bibitem[{\citenamefont{Garz\'o}(2017)}]{G17}
\bibinfo{author}{\bibnamefont{Garz\'o}} V, 2017
  \bibinfo{journal}{Phys. Rev. E} \textbf{\bibinfo{volume}{95}},
  \bibinfo{pages}{062906}

\bibitem[{\citenamefont{Goldhirsch}(2003)}]{G03}
\bibinfo{author}{\bibnamefont{Goldhirsch} I}, 2003
  \bibinfo{journal}{Annu. Rev. Fluid Mech.} \textbf{\bibinfo{volume}{35}},
  \bibinfo{pages}{267}

\bibitem[{\citenamefont{Garz\'o and Santos}(1995)}]{GS95a}
\bibinfo{author}{\bibnamefont{Garz\'o} V} \bibnamefont{and}
  \bibinfo{author}{\bibnamefont{Santos} A}, 1995
  \bibinfo{journal}{Physica A} \textbf{\bibinfo{volume}{213}},
  \bibinfo{pages}{426}

\bibitem[{\citenamefont{Bird}(1994)}]{B94}
\bibinfo{author}{\bibnamefont{Bird} G A}, 1994
  \emph{\bibinfo{title}{Molecular Gas Dynamics and the Direct Simulation Monte
  Carlo of Gas Flows}} (\bibinfo{publisher}{Clarendon: Oxford},


\bibitem[{\citenamefont{Bridges et~al.}(1984)\citenamefont{Bridges, Hatzes, and
  Lin}}]{BHL84}
\bibinfo{author}{ \bibnamefont{Bridges} F G},
  \bibinfo{author}{\bibnamefont{Hatzes} A} \bibnamefont{and}
  \bibinfo{author}{\bibnamefont{Lin} D N C}, 1984
  \bibinfo{journal}{Nature (London)} \textbf{\bibinfo{volume}{309}},
  \bibinfo{pages}{333}

\bibitem[{\citenamefont{Brilliantov and P\"oschel}(2000)}]{BP00}
\bibinfo{author}{\bibnamefont{Brilliantov} N V}
  \bibnamefont{and}
  \bibinfo{author}{\bibnamefont{P\"oschel} T}, 2000
  \bibinfo{journal}{Phys. Rev. E} \textbf{\bibinfo{volume}{61}},
  \bibinfo{pages}{5573}

\bibitem[{\citenamefont{Brilliantov and P\"oschel}(2003)}]{BP03}
\bibinfo{author}{\bibnamefont{Brilliantov} N V}
  \bibnamefont{and}
  \bibinfo{author}{\bibnamefont{P\"oschel} T}, 2003
  \bibinfo{journal}{Phys. Rev. E} \textbf{\bibinfo{volume}{67}},
  \bibinfo{pages}{{061}{304}}

\bibitem[{\citenamefont{Dubey et~al.}(2013)\citenamefont{Dubey, Bodrova, Puri,
  and Brilliantov}}]{DBPB13}
\bibinfo{author}{\bibnamefont{Dubey} A K},
  \bibinfo{author}{\bibnamefont{Bodrova} A},
  \bibinfo{author}{\bibnamefont{Puri} S} \bibnamefont{and}
  \bibinfo{author}{\bibnamefont{Brilliantov} N V}, 2013
  \bibinfo{journal}{Phys. Rev. E} \textbf{\bibinfo{volume}{87}},
  \bibinfo{pages}{{062}{202}}


\bibitem{GS95}Goldshtein A and Shapiro M, 1995 J. Fluid Mech. \textbf{282}, 75 

\bibitem{Z06}Zippelius A, 2006 Physica A \textbf{369}, 143


\end{thebibliography}

\end{document}